\documentclass[twocolumn,numberedappendix,iop]{openjournal}
\usepackage{lineno}
\usepackage{graphicx,amsmath,amssymb,amstext}
\usepackage{amsbsy,amsfonts,amsthm,color}
\usepackage[colorlinks,linkcolor=blue,citecolor=blue,urlcolor=blue]{hyperref}
\usepackage[utf8]{inputenc}
\usepackage{float}
\usepackage{xcolor}
\usepackage{ulem}
\usepackage[T1]{fontenc}
\usepackage[title]{appendix}
\usepackage{savesym}
\savesymbol{tablenum}
\usepackage{siunitx}
\restoresymbol{SIX}{tablenum}
\DeclareSIUnit\h{\text{$h$}}

\begin{document}
\title{Detection of supernova magnitude fluctuations induced by large-scale structure \vspace{-4em}}

\author{
A.~Nguyen,$^{1,2,*}$
C.~Blake,$^{1,2}$
R. J.~Turner,$^{1,2}$
V.~Aronica,$^{3}$
J.~Bautista,$^{3}$
J.~Aguilar,$^{4}$
S.~Ahlen,$^{5}$
S.~BenZvi,$^{6}$
D.~Bianchi,$^{7,8}$
D.~Brooks,$^{9}$
A.~Carr,$^{10}$
T.~Claybaugh,$^{4}$
A.~Cuceu,$^{4}$
A.~de la Macorra,$^{11}$
B.~Dey,$^{12,13}$
P.~Doel,$^{9}$
K.~Douglass,$^{6}$
S.~Ferraro,$^{4,14}$
J.~E.~Forero-Romero,$^{15,16}$
E.~Gaztañaga,$^{17,18,19}$
S.~Gontcho A Gontcho,$^{4,20}$
G.~Gutierrez,$^{21}$
J.~Guy,$^{4}$
K.~Honscheid,$^{22,23,24}$
C.~Howlett,$^{25}$
D.~Huterer,$^{26,27}$
M.~Ishak,$^{28}$
R.~Joyce,$^{29}$
R.~Kehoe,$^{30}$
A.~G.~Kim,$^{4}$
A.~Kremin,$^{4}$
O.~Lahav,$^{9}$
M.~Landriau,$^{4}$
L.~Le~Guillou,$^{31}$
A.~Leauthaud,$^{32,33}$
M.~E.~Levi,$^{4}$
M.~Manera,$^{34,35}$
P.~Martini,$^{22,36,24}$
A.~Meisner,$^{29}$
R.~Miquel,$^{37,35}$
E.~Mueller,$^{38}$
S.~Nadathur,$^{18}$
N.~Palanque-Delabrouille,$^{39,4}$
W.~J.~Percival,$^{40,41,42}$
C.~Poppett,$^{4,43,14}$
F.~Prada,$^{44}$
F.~Qin,$^{3}$
A.~J.~Ross,$^{22,36,24}$
C.~Ross,$^{25}$
G.~Rossi,$^{45}$
E.~Sanchez,$^{46}$
D.~Schlegel,$^{4}$
M.~Schubnell,$^{26,27}$
D.~Sprayberry,$^{29}$
G.~Tarl\'{e},$^{27}$
B.~A.~Weaver,$^{29}$
P.~Zarrouk,$^{31}$
R.~Zhou,$^{4}$
and H.~Zou$^{47}$ \\
{\it (Affiliations can be found after the references)}
}
\thanks{$^*$E-mail: andnguyen@swin.edu.au}

\begin{abstract}
The peculiar velocities of supernovae and their host galaxies are correlated with the large-scale structure of the Universe, and can be used to constrain the growth rate of structure and test the cosmological model.  In this work, we measure the correlation statistics of the large-scale structure traced by the Dark Energy Spectroscopic Instrument Bright Galaxy Survey Data Release 1 sample, and magnitude fluctuations of type Ia supernova from the Pantheon+ compilation across redshifts $z < 0.1$.  We find a detection of the cross-correlation signal between galaxies and type Ia supernova magnitudes. Fitting the normalised growth rate of structure $f \sigma_8$ to the auto- and cross-correlation function measurements we find $f \sigma_8 = 0.384^{+0.094}_{-0.157}$ at $z = 0.03$, which is consistent with the {\it Planck} $\Lambda$CDM model prediction, and indicates that the supernova magnitude fluctuations are induced by peculiar velocities. Using a large ensemble of N-body simulations, we validate our methodology, calibrate the covariance of the measurements, and demonstrate that our results are insensitive to supernova selection effects. We highlight the potential of this methodology for measuring the growth rate of structure, and forecast that the next generation of type Ia supernova surveys will improve $f \sigma_8$ constraints by a further order of magnitude.
  \\[1em]
\end{abstract}

\maketitle

\section{Introduction}
\label{sec:intro}

Since the discovery of the accelerating expansion of the Universe \citep{riess1998, schmidt1998, perlmutter1999}, type Ia supernovae (SNe Ia), as standardisable candles, have served as a leading probe of the cosmic expansion history. Fits to SN Ia photometric light curves provide their apparent magnitude, stretch and colour, which are used to standardise their peak brightnesses. Together with the redshifts of the SN host galaxy from spectroscopic observations, we obtain the magnitude-redshift relation, from which we can infer the expansion rate of the Universe \citep[for recent cosmological studies see,][]{2022ApJ...934L...7R, 2022ApJ...938..110B, 2025ApJ...986..231R, 2024ApJ...973L..14D}.

The majority of SN Ia cosmological studies have focused on improving photometric calibration and other systematics to increase accuracy and precision of peak brightness measurements, enabling more accurate inference of the Hubble constant. Typically, the scatter about the underlying magnitude-redshift relation is treated as an uncertainty and propagated into the error budget \citep{davis2011b}. Whilst this scatter does include a component originating from the intrinsic brightness variation of SNe Ia, it also contains coherent fluctuations due to the peculiar velocities (PVs) of the SN Ia host galaxies. The dominant contribution to these fluctuations is from peculiar velocities at low redshift, so we focus on redshift range $z < 0.1$ in this study. There is also a contribution due to gravitational lensing, however this effect is only perceptible in higher-redshift SNe and was not considered in this study.

The peculiar velocity of a galaxy is its motion relative to the Hubble rest frame. Gravitational instability leads to overdensities whose gravitational attraction influences galaxy motion. Peculiar velocities are therefore correlated with the galaxy density field and provide a test of the cosmological model \citep{strauss1995, davis2011b}. The velocity field is more sensitive than the density field to matter fluctuations on the largest scales \citep[e.g.][]{2014MNRAS.445.4267K}, making it a powerful probe of the underlying matter distribution, gravity and dark energy. Velocity fluctuations can be converted to magnitude fluctuations \citep{hui2006, amendola2021} and compared with correlations observed in SN datasets \citep[e.g.,][]{1997ApJ...488L...1R, 2007PhRvL..99h1301G, johnson2014, 2016PDU....13...66C, huterer2017, 2023A&A...674A.197C}.  In a complementary approach, \citet{2017MNRAS.467..259M} constrained the cosmological model by studying the moments of the supernova magnitude distribution.  Similar information is imprinted in the cross-correlation between SN magnitude fluctuations and large-scale structure, which we consider in this study.

The $\Lambda$CDM model of cosmology describes a Universe dominated by dark energy $\Lambda$ and cold dark matter (CDM), in which gravity is characterised by general relativity at all scales. However, the lack of a theoretical justification for dark energy continues to motivate alternative cosmological models. The growth rate of structure $f$ is determined by the evolution of density perturbations under the influence of gravity. Peculiar velocities constrain a degenerate combination of $f$, the growth rate of structure, and $\sigma_8$, the amplitude of density fluctuations on scales of $8 \, h^{-1}$ Mpc, serving as a probe of the underlying gravitational physics and offering constraints on modified gravity theories \citep[for a recent review, see][]{2024arXiv241119484T}. Because of this degeneracy it is common for peculiar velocity measurements to constrain a combined parameter referred to as the normalised growth rate of structure, $f \sigma_8$.

Peculiar velocity correlations have been previously measured by a range of ``standard candle'' probes and methodologies.  SNe Ia, as standardisable candles, enable more accurate inference of peculiar velocities because of their smaller intrinsic scatter in standardised peak brightness in comparison to other independent distance indicators such as the Tully-Fisher \citep{tully1977} and Fundamental Plane \citep{djorgovski1987, dressler1987} methods. However, due to the historically limited sample size and sky coverage of existing SN Ia surveys, and the inhomogeneous sky coverage of SN Ia compilations, inference of $f\sigma_8$ using SNe Ia \citep[e.g.,][]{johnson2014, huterer2017, boruah2020} has been less extensive when compared to the other distance indicators. 

Methods for constraining $f \sigma_8$ using peculiar velocities include: the two-point correlation function, \citep{nusser2017,2017PhRvD..95h3502A, 2018MNRAS.480.5332W, dupuy2019, 2023MNRAS.518.2436T,courtois2023, lyall2024}; the maximum-likelihood fields method, where the covariance between the peculiar velocity and galaxy overdensity fields is modelled analytically \citep{johnson2014, huterer2017, adams2017, adams2020, howlett2017b, lai2023}; measuring the momentum power spectrum \citep{park2000, park2006, howlett2019, qin2019, 2025ApJ...978....7Q}; and reconstructions of the peculiar velocity field \citep{davis2011a, carrick2015, boruah2020, said2020, lilow2021, 2023JCAP...06..062Q, boubel2024, 2024MNRAS.531..788H}. As a whole, these methods have produced growth rate measurements that are largely in agreement with the predictions of the $\Lambda$CDM cosmology, though with significant $\sim 20\%$ errors in most cases. Given the current limited samples of SNe Ia, it is not expected they will currently have constraining power competitive with the methods outlined above, but these constraints will continue to improve as new SNe Ia are detected.

The goal of this paper is to perform new $f \sigma_8$ measurements using the correlations between SN Ia magnitudes and large-scale structure. We use SNe Ia from the Pantheon+ compilation \citep{2022ApJ...938..113S}, one of the largest samples of SNe Ia currently available, and for our large-scale structure maps, we use the Bright Galaxy Survey (BGS) component of Data Release 1 (DR1) of the Dark Energy Spectroscopic Instrument \citep[DESI,][]{2016arXiv161100036D, 2023AJ....165..253H}. DESI is a galaxy-redshift survey being conducted on the 4m Mayall Telescope, at the Kitt Peak National Observatory.  The DESI BGS offers an order-of-magnitude increase in the number of large-scale structure tracers over the previous generation of surveys, resulting in an increase in the significance of the potential correlation signal.  This is the first study using the DESI density field in a cross-correlation analysis with SNe Ia.  We measure the correlation statistics between these datasets using estimators of the magnitude auto-correlation functions $\psi_1$ and $\psi_2$ \citep{gorski1988}, and the galaxy-magnitude cross-correlation function $\psi_3$ \citep{turner2021}, and test $\Lambda$CDM by fitting $f\sigma_8$ to these measurements. We validate our models using simulations, and forecast how future large-scale structure and SN Ia surveys could improve our constraints on $f\sigma_8$.

With new generation surveys such as the Zwicky Transient Facility \citep[ZTF,][]{2019PASP..131a8002B, 2025A&A...694A...1R} and the Rubin Observatory Legacy Survey of Space and Time \citep[Rubin-LSST,][]{2019ApJ...873..111I} currently on-sky or in development, we will observe future SN samples with high cadence and large sky coverage. From this, a more comprehensive and uniform sample of SNe Ia will soon be available for peculiar velocity studies \citep{howlett2017, 2023A&A...674A.197C}. Local large-scale structure datasets offer direct peculiar velocity measurements and the DESI PV survey plans to measure the peculiar velocities of 186,000 galaxies over its 5 years of operation \citep{2023MNRAS.525.1106S}. This will be further expanded by surveys such as the 4-metre Multi-Object Spectroscopic Telescope \citep[4MOST;][]{dejong2019} Hemisphere Survey \citep[4HS,][]{taylor2023} and the Australian Square Kilometre Array Pathfinder WALLABY survey \citep{2020Ap&SS.365..118K}. The influx of data from these surveys will improve our distance estimates and combining SN Ia, Tully-Fisher and Fundamental Plane peculiar velocity measurements will provide tighter constraints on $f\sigma_8$ and, consequently, modified gravity models \citep{kim2020, lyall2022}.

Our paper is structured as follows.  The theoretical models we use to describe the magnitude fluctuations induced by peculiar velocities are outlined in Sec.~\ref{sec:theory}, and the datasets we utilise from DESI and Pantheon+ are described in Sec.~\ref{sec:data}.  The N-body simulations we created to test the theoretical models and the impact of selection effects on our analysis are discussed in Sec.~\ref{sec:abacus}, and the magnitude auto-correlation functions $\psi_1$ and $\psi_2$, galaxy-magnitude cross-correlation function $\psi_3$, and galaxy auto-correlation function $\xi_{gg}$ are measured in Sec.~\ref{sec:measurements}. The growth rate analysis validation using simulations is shown in Sec.~\ref{sec:validation}, and the $f\sigma_8$ constraint is presented in Sec.~\ref{sec:fits}.  We summarise our conclusions and forecast the impact of new large-scale structure and SN Ia surveys in Sec.~\ref{sec:disc}.

\section{Magnitude Fluctuation Correlation Theory}
\label{sec:theory}

\subsection{Growth rate of structure}

The growth rate of structure $f$ describes the evolution of density perturbations under the influence of gravity. In the linear theory of structure formation, density fluctuations $\delta$ grow as $\delta \propto D(a)$, where $D(a)$ is the linear growth factor, dependent on cosmic scale factor $a$.  The growth rate is then defined as
\begin{equation}
    f(a) = \frac{d\ln{D(a)}}{d\ln{a}}.
\end{equation}

Measuring $f$ allows us to test gravitational physics and constrain modified gravity and dark energy models. The growth rate at redshift $z$ can be parameterised as 
\begin{equation}
    f(z) = \Omega_m(z)^{\gamma},
\end{equation}
where $\Omega_m$ is the matter density parameter and $\gamma$ is the growth index, as shown by \citet{1998ApJ...508..483W}. In the $\Lambda$CDM model, $\gamma = {0.55}$ \citep{linder2007}.  The value of $\gamma$ varies in different gravity and dark energy models. For the Dvali, Gabadadze $\&$ Porrati braneworld model \citep{2000PhLB..485..208D}, $\gamma = 0.6875$ \citep{linder2007}, and in $f(R)$ gravity models, which include a non-linear function of the Ricci scalar $R$ in the Einstein-Hilbert action, the growth rate is dependent on scale \citep{2007PhRvD..76f4004H}. The growth rate modulates the relation between the local peculiar velocity $\mathbf{v}(\mathbf{x})$ and matter overdensity $\delta_m(\mathbf{x})$, at some point $\mathbf{x}$, via the linear theory continuity equation,

\begin{equation}
\mathbf{\nabla} \cdot \mathbf{v}(\mathbf{x}) = -a H f \, \delta_m(\mathbf{x}),
\end{equation}
where $H$ is the Hubble parameter.

\subsection{Relation between magnitude and velocity fluctuations}
\label{sec:veltomag}

In this section, we derive the relation for the magnitude fluctuation imprinted by the velocity of a source \citep[following][]{hui2006,amendola2021}.  We start from the relation between apparent magnitude $m$ and luminosity distance $D_L$ in \unit{Mpc}, given by the distance modulus equation,
\begin{equation}
    m = M + 5 \log_{10}{D_L} + 25,
\label{eq:dm}
\end{equation}
where $M$ is the absolute magnitude.  From Eq.~\ref{eq:dm}, we can show that the relation between fluctuations in magnitude $\delta m$ and luminosity distance $\delta D_L$ is,
\begin{equation}
    \delta m = \frac{5}{\ln{10}} \frac{\delta D_L}{D_L} .
\label{eq:dmdl}
\end{equation}
The luminosity distance fluctuations $\delta D_L$ can be determined by relating observations in a perturbed universe to those in a homogeneous Universe (for which quantities will be denoted with an overline).  The perturbation in redshift is given by,
\begin{equation}
    1 + z = (1 + \overline{z}) \left( 1 + \frac{v_r}{c} \right),
\end{equation}
where $v_r$ is the radial velocity of the source.  The relation between the luminosity distance $D_L$ and angular diameter distance $D_A$ applies in both perturbed and unperturbed Universes \citep{hui2006, 2019MNRAS.490.2948D},
\begin{equation}
\begin{split}
    D_L(z) &= D_A(z) (1 + z)^2 , \\
    \overline{D}_L(\overline{z}) &= \overline{D}_A(\overline{z}) (1 + \overline{z})^2.
\end{split}
\end{equation}
Neglecting observer motion, we can write $D_A(z) = \overline{D}_A(\overline{z})$, hence,
\begin{equation}
    \frac{D_L(z)}{\overline{D}_L(\overline{z})} = \frac{(1+z)^2}{(1+\overline{z})^2} = \left( 1 + \frac{v_r}{c} \right)^2 \approx 1 + \frac{2v_r}{c}.
\label{eq:dl2}
\end{equation}
We now perform a Taylor series expansion about $z = \overline{z}$,
\begin{equation}
    \overline{D}_L(z) = \overline{D}_L(\overline{z}) + \frac{\partial \overline{D}_L}{\partial z} \big\vert_{z = \overline{z}} \left( z - \overline{z} \right) .
\end{equation}
Using $\overline{D}_L(z) = \chi(z) \, (1+z)$ and $d\chi/dz = c/H(z)$ in terms of radial co-moving co-ordinate $\chi(z)$, we find,
\begin{equation}
    \overline{D}_L(z) = \overline{D}_L(\overline{z}) \left\{ 1 + \left[ 1 + \frac{c \, (1 + \overline{z})^2}{H(\overline{z}) \, D_L(\overline{z})} \right] \frac{v_r}{c} \right\} .
\label{eq:dl3}
\end{equation}
Substituting Eq.~\ref{eq:dl2} and Eq.~\ref{eq:dl3} into Eq.~\ref{eq:dmdl} produces the magnitude fluctuation $\delta m$ in terms of the radial velocity, which can be conveniently written,
\begin{equation}
  \delta m = \alpha(z) \times v_r ,
\label{eq:alpha}
\end{equation}
where $\alpha$ is the coefficient of proportionality between velocity and magnitude,
\begin{equation}
    \alpha(z) = \frac{5}{c \, \ln{10}} \left[ 1 - \frac{c \, (1+z)^2}{H(z) \, D_L(z)} \right] .
\label{eq:veltomag}
\end{equation}
Using Eq.~\ref{eq:alpha}, we can relate the measurement of magnitude fluctuation correlations to velocity correlations.

\subsection{Galaxy and velocity correlation functions}
\label{sec:model}

As shown in Sec.~\ref{sec:veltomag}, magnitude fluctuations can be generated by the peculiar motions of galaxies.  By modelling the velocity correlations induced by the growth of large-scale structure, we can predict the observed correlation spectrum of magnitudes.  For a Gaussian peculiar velocity field, the two-point correlation tensor between two positions $A$ and $B$ is given by \citet{gorski1988, gorski1989} as,
\begin{equation}
\label{eq:corrtensor}
    \Psi_{ij}(\mathbf{r}_A, \mathbf{r}_B) = \langle v_i(\mathbf{r}_A) \, v_j(\mathbf{r}_B) \rangle,
\end{equation}
where $\mathbf{r}$ is a spatial position and $v_i$ is the peculiar velocity component in directions $i = \{ x, y, z \}$. For an irrotational, homogeneous and isotropic velocity field with linear velocity perturbations, the velocity correlation tensor can be written as \citep[e.g.,][]{gorski1988, gorski1989, 2024MNRAS.527..501B},
\begin{equation}
    \Psi_{ij}(r) = \left[\Psi_{\parallel}(r) - \Psi_{\perp}(r)\right] \, \hat{r}_{Ai} \, \hat{r}_{Bj} + \Psi_{\perp}(r) \, \delta^K_{ij},
\end{equation}
where $r$ is the separation between positions $A$ and $B$, $\Psi_{\parallel}(r)$ and $\Psi_{\perp}(r)$ are the correlation functions between components of velocity parallel and perpendicular to the separation vector $\mathbf{r}$, and $\delta^K_{ij}$ is the Kronecker delta. 

The correlation functions $\Psi_{\parallel}$ and $\Psi_{\perp}$ can be written in terms of the matter power spectrum $P(k)$ as,
\begin{equation}
\label{eq:corrperppar}
\begin{split}
    \Psi_{\parallel}(r) &= \frac{H^2 a^2 (f\sigma_8)^2}{2\pi^2} \int \frac{P(k)}{\sigma_{8,\mathrm{fid}}^2} \left[j_0(kr) - 2\frac{j_1(kr)}{kr}\right] \, dk, \\
    \Psi_{\perp}(r) &= \frac{H^2 a^2 (f\sigma_8)^2}{2\pi^2} \int \frac{P(k)}{\sigma_{8,\mathrm{fid}}^2} \frac{j_1(kr)}{kr} \, dk,
\end{split}
\end{equation}
where $P(k)$ is the model matter power spectrum as a function of wavenumber $k$, $\sigma_{8,\mathrm{fid}}$ is the chosen normalisation of the matter power spectrum, and $j_\ell$ are spherical Bessel functions of the first kind,
\begin{equation}
\begin{split}
    j_0(x) &= \frac{\sin(x)}{x}, \\
    j_1(x) &= \frac{\sin(x)}{x^2} - \frac{\cos(x)}{x}.
\end{split}
\end{equation}
The formulation of Eq.~\ref{eq:corrperppar} illustrates that the measured velocity correlation function may be used to constrain $f\sigma_8$. 

Peculiar velocities are also correlated with galaxy positions, where the cross-correlation function is given by \citep[e.g.,][]{adams2017},
\begin{equation}
    \xi_{gv}(r) = -\frac{H a (f \sigma_8) (b \sigma_8)}{2\pi^2} \int dk \, k \, \frac{P(k)}{\sigma_{8,\mathrm{fid}}^2} \, j_1(kr) ,
\label{eq:galvelcorr}
\end{equation}
where $b$ is the linear galaxy bias describing how galaxy distribution traces the underlying matter density field, and $b\sigma_8$ is the the normalised linear galaxy bias. Finally, the galaxy auto-correlation function monopole model (neglecting redshift-space distortions for the moment) is,
\begin{equation}
    \xi_{gg}(r) = \frac{(b \sigma_8)^2}{2\pi^2} \int dk \, k^2 \, \frac{P(k)}{\sigma_{8,\mathrm{fid}}^2} \, j_0(kr) .
\label{eq:galcorr}
\end{equation}
We generate the linear model power spectrum $P(k)$ using \texttt{CAMB} \citep{lewis2000}, and use the halofit non-linear matter power spectrum \citep{takahashi2012} to model the velocity and galaxy correlation functions.

Since we wish to focus on the growth information contained in SN magnitude correlations in this study, we do not consider the galaxy auto-correlation quadrupole, which adds significant additional information through the effect of redshift-space distortions (RSD).  We include the effect of RSD on the remaining statistics by scaling the galaxy auto-correlation monopole by a linear RSD factor $(1 + \frac{2}{3}\beta + \frac{1}{5} \beta^2)$, where $\beta = f/b$, and the galaxy-velocity cross-correlation function by $(1 + \frac{1}{3}\beta)$ \citep[e.g.][]{2023MNRAS.518.2436T}.  As we are restricting our model fits to large scales, we find that linear bias and RSD models are acceptable, which we test using mock catalogues in Sec.~\ref{sec:validation}.

\section{Density field and magnitude fluctuation datasets}
\label{sec:data}

\subsection{The Dark Energy Spectroscopic Instrument}

DESI is using the 4m Mayall Telescope, at the Kitt Peak National Observatory, to perform a spectroscopic survey targeting galaxies and quasars to measure the expansion history of the Universe and the growth of large-scale structure \citep{2013arXiv1308.0847L, 2016arXiv161100036D, 2016arXiv161100037D, 2025JCAP...07..028A}.  The DESI Survey is planned to span five years, during which it will observe more than 40 million galaxies and quasars over a footprint of 14,000 square degrees.  The survey is divided into four main target classes: Bright Galaxy Survey (BGS), Luminous Red Galaxy Survey (LRG), Emission Line Galaxy Survey (ELG), and Quasar Survey (QSO), which are selected from the DESI Legacy optical imaging surveys \citep{2019AJ....157..168D}.  Targets within each observational field are assigned to 5000 optical fibres \citep{2024AJ....168..245P} in the telescope focal plane using a robotic positioner \citep{2023AJ....165....9S} and optical corrector \citep{2024AJ....168...95M}, and the observed data are processed by the DESI spectroscopic pipeline \citep{2023AJ....165..144G}.  The overall observing strategy is summarised by \cite{2023AJ....166..259S}.  DESI has issued an Early Data Release \citep{2024AJ....168...58D}, which has been used for scientific validation of the program \citep{2024AJ....167...62D}. In this study, we use a galaxy sample from the publicly available DESI Data Release 1 \citep[DR1;][]{2025arXiv250314745D}, comprising of spectra obtained during the first year of full-survey observations.

\subsection{The Bright Galaxy Survey dataset}

We draw our large-scale structure data from the DESI Bright Galaxy Survey (BGS), which targets a magnitude-limited sample of galaxies with $r$-band magnitudes $14 < r < 19.5$ across redshifts $z < 0.5$.  BGS targets are assigned high priority during DESI bright-time observations.  The full selection criteria and validation of the BGS sample are described by \cite{2023AJ....165..253H}, resulting in a sample with density 854 deg$^{-2}$ containing reliable redshift measurements for over 5.5 million galaxies in DR1.  The BGS sample consists of a magnitude-limited Bright sample with $r < 19.5$, and a colour-selected Faint component with $19.5 < r < 20.175$. For this study, we will only consider the BGS Bright sample, as the Faint sample suffers from complications regarding incompleteness and systematics \citep{2023AJ....165..253H}.

Large-scale structure catalogues suitable for cosmological analysis have been built from the redshift and target catalogues for each DESI tracer as described by \cite{2025JCAP...01..125R}.  The selection function of the sample is defined, along with correction weights designed to compensate for systematic density variations with observational characteristics across the survey.  These selection functions yield accompanying unclustered random catalogues which are used in correlation function estimators.  Measurements of Baryon Acoustic Oscillations in the clustering pattern of the DR1 BGS sample are presented by \cite{2025JCAP...04..012A}.

\begin{figure*}
    \centering
    \includegraphics[width=\textwidth]{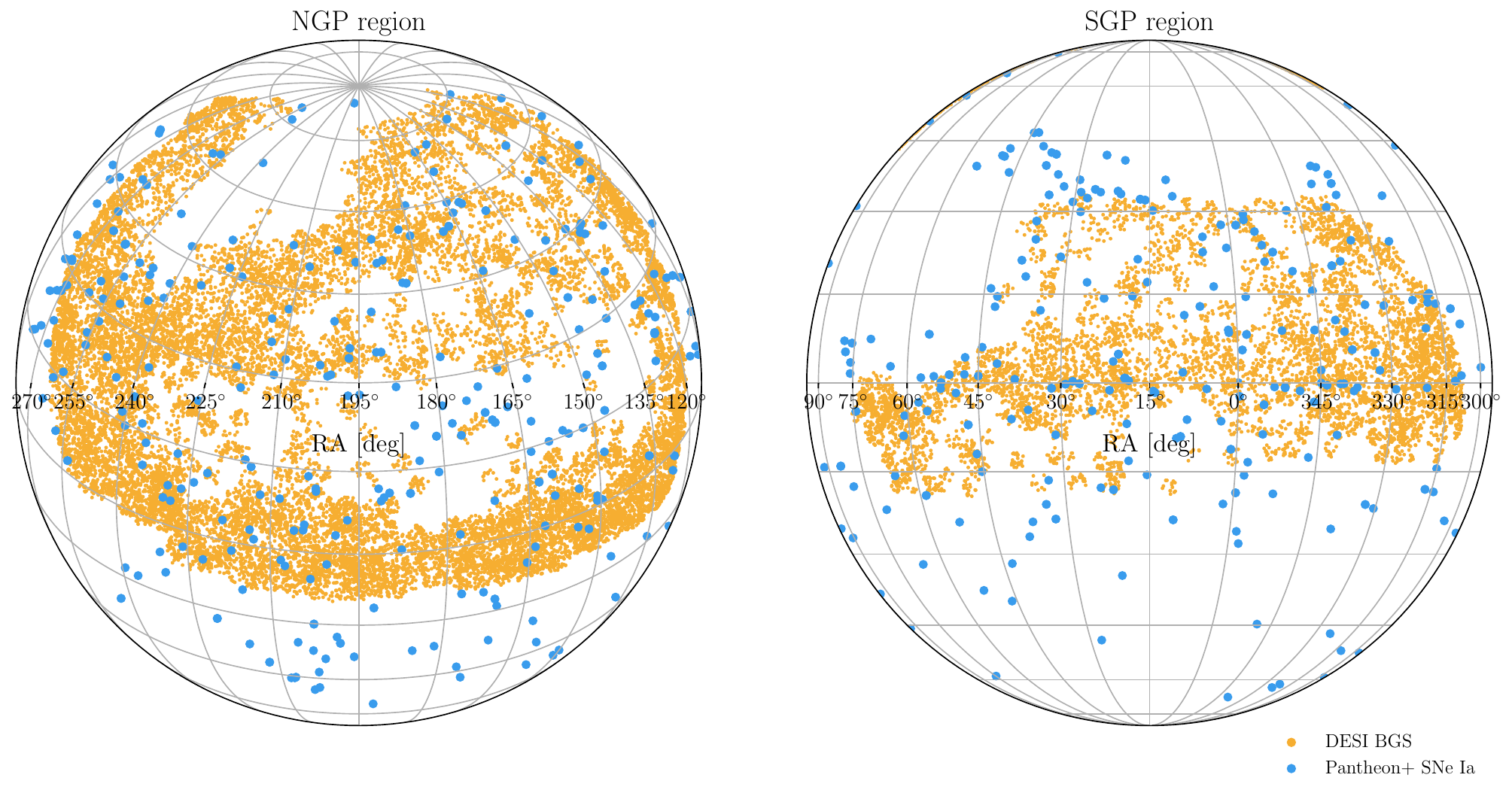}
    \caption{Sky map showing the (RA,Dec) distribution of DESI BGS galaxies and Pantheon+ SNe at $z < 0.1$, centred on the North Galactic Pole (NGP) and South Galactic Pole (SGP) regions.}
\label{fig:skymap}
\end{figure*}

\begin{figure}
    \includegraphics[width=\columnwidth]{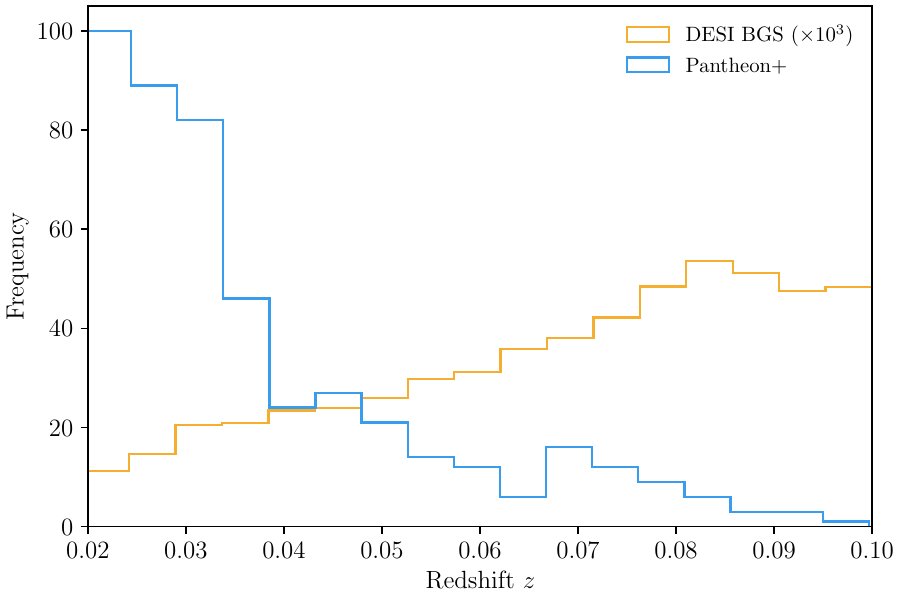}
    \caption{The normalised redshift probability distributions of DESI BGS galaxies and Pantheon+ SNe at $z < 0.1$.  For clarity of presentation, we have applied smoothing to the raw redshift histograms.}
\label{fig:reddist}
\end{figure}

To create a sub-sample overlapping with low-redshift supernovae, we restrict our analysis to the redshift range $z < 0.1$, resulting in a sample of 578,576 galaxies. The sky positions of the BGS galaxies are shown in Fig.~\ref{fig:skymap} and the redshift distribution of the sample is shown in Fig.~\ref{fig:reddist}.

\subsection{Pantheon+ dataset}

The Pantheon+ dataset contains 1701 light curves of 1550 distinct spectroscopically-confirmed type Ia supernovae (SNe Ia) in redshift range $0.001 < z < 2.261$ \citep{2022ApJ...938..113S}. A compilation of 18 supernova surveys, the Pantheon+ catalogue standardises SN Ia light curves using a consistent magnitude cross-calibration \citep{2022ApJ...938..111B}. The distance modulus $\mu$ for a SN Ia is determined by fitting the light curves with the SALT2 model originally developed by \cite{2010A&A...523A...7G} and updated in \cite{2022ApJ...938..111B}.  This results in measurements of the peak magnitude or light curve amplitude $m_B$, the stretch parameter $x_1$ and light-curve color $c$. These light-curve fit parameters are related to the distance modulus using the modified Tripp relation \citep{1998A&A...331..815T, 2017ApJ...836...56K}:
\begin{equation}
    \mu = m_B + \alpha x_1 - \beta c - M - \delta_{\mu \mathrm{-bias}} ,
\label{eq:distmod}
\end{equation}
where the correlation coefficients $\alpha = 0.14$ and $\beta=3.1$, $M$ is the fiducial SN Ia absolute magnitude and $\delta_{\mu \mathrm{-bias}}$ is the bias correction derived from simulations which account for selection effects and systematics involved in distance recovery \citep{2021ApJ...913...49P}. Cosmological constraints from the Pantheon+ dataset are presented by \cite{2022ApJ...938..110B}.

As magnitude fluctuations are primarily due to peculiar velocities at low redshift, we restrict the SN sample to $z < 0.1$, which provides 510 unique SNe Ia. The effective redshift of the measurement of $f\sigma_8$ is $z_\mathrm{eff} = 0.03$, representing the mean redshift of the SN sample. The Pantheon+ SN Ia sky positions and redshift distributions are shown in Fig.~\ref{fig:skymap} and Fig.~\ref{fig:reddist}, respectively.  From the Pantheon+ SN Ia dataset\footnote{Accessed from \url{https://github.com/PantheonPlusSH0ES/DataRelease/tree/main/Pantheon\%2B_Data/4\_DISTANCES\_AND\_COVAR}.} we used the CMB-frame redshift, right ascension, declination, magnitude, and magnitude error of each SN Ia. We determined the magnitude fluctuation $\delta m$ in each case by subtracting the fiducial magnitude-redshift relation assuming a flat $\Lambda$CDM Universe with $H_0 = 100 \, h$~\unit{km.s^{-1}.Mpc^{-1}}, $\Omega_m = 0.3153$, $\Omega_{\Lambda} = 0.6847$, where the density parameters represent the DESI fiducial cosmology, but do not have a significant effect for our low-redshift sample.  We subtract the ensemble mean from the $\delta m$ values (effectively fitting for the absolute magnitude $M - 5 \log_{10} h$).

\section{Abacus N-body Simulations}
\label{sec:abacus}

In this section we summarise the N-body simulation datasets we used to test and validate our methodology, theoretical models, and the impact of supernovae selection effects on our analysis. We do this by utilising mocks matched to the DESI BGS sample, including a realistic population of SN host galaxies.  These mocks include realistic tracer selection functions, measurement errors and galaxy bias.  We constructed these mock catalogues from the $z=0.2$ snapshot of the AbacusSummit suite of N-body simulations \citep{garrisonABACUSCosmologicalNbody2021, maksimovaABACUSSUMMITMassiveSet2021, hadzhiyskaCOMPASONewHalo2022}, the lowest-redshift output, which includes 25 independent boxes generated in the {\it Planck 2018} best-fit flat $\Lambda$CDM cosmology using fiducial parameters, $\Omega_m = 0.3153$, $\Omega_{\Lambda} = 0.6847$, $\Omega_b = 0.0493$, $h = 0.6736$, $\sigma_8 = 0.8114$ and $n_s = 0.9649$.

In order to simulate the DESI BGS distribution, dark matter halos were populated with a luminosity-dependent halo occupation distribution model \citep{smithLightconeCatalogueMillenniumXXL2017, smithLightconeCatalogueMillenniumXXL2022}, reproducing the comoving number density of galaxies as well as the projected correlation function of DESI BGS data \citep{2023AJ....165..253H}.  In addition to positions and velocities, mock galaxies have observed magnitudes and colours that follow a realistic luminosity function including redshift evolution.  The full details of the mock production can be found in \citet{smithGeneratingMockGalaxy2024} and references therein. 

As we are conducting an analysis of the low-redshift Universe ($z<0.1$), we divided the initial 2 $h^{-1}$~Gpc simulation boxes into $3^3$ sub-volumes, placing the observer at the centre of each.  Given that we have 25 independent Abacus boxes, this yields a total of 675 mock catalogues.  While peculiar velocites are correlated on large scales, we assume all sub-volumes to be statistically independent,  which is a good approximation for the purposes of this work.  We apply observational completeness masks corresponding to the DESI DR1 dataset.  Additional properties, such as stellar masses, star formation rates and galaxy sizes are assigned from real data to mock galaxies based on proximity in redshift-magnitude-colour space.

We populated our mock BGS galaxies with SNe Ia that have explosion rates $R_{Ia}$ dependent on stellar mass $M_*$ and star-formation rate (SFR), known as the ``A+B model'' \citep{mannucci_supernova_2005}:
\begin{equation}
\label{eq:snIa rate}
    R_{Ia} = A \times M_*^{n_M} + B \times \mathrm{SFR}^{n_S} ,
\end{equation}
where we assume a linear relation ($n_M = n_S = 1$), $A = 5.3 \times 10^{-14}$ \si{yr^{-1}.M_\odot^{-1}} and $B = 3.9 \times 10^{-4}$ \si{yr^{-1}. (M_\odot.yr^{-1})^{-1}} \citep{sullivan_rates_2006}. For a given galaxy, the number of SN Ia is a Poisson realisation of the rate multiplied by the number of years of observations.  The model for the supernova rate consists of a ``delayed'' component, with a long delay time driven  by the stellar mass of the galaxy, and a ``prompt'' component, with short delay times caused by the formation of new stars.

\subsection{Modelling SN properties for selection effects}
\label{sec:seleff}

Supernova samples contain observational selection effects which may imprint magnitude- or redshift-dependent trends, presenting potential systematics in the interpretation of magnitude fluctuations as tracing peculiar velocities.  In this section we describe how we apply these effects to our simulated Abacus samples to assess their influence on growth rate constraints. We test the influence of SN selection effects by performing growth rate fits with and without applying a range of magnitude selections to our simulated samples. Using Eq.~\ref{eq:distmod}, we model the rest-frame absolute magnitude of each SN Ia (labelled $i$) as,
\begin{equation}
    M^*_i = M_B - \alpha x_{1,i} + \beta c_i + \sigma_{\mathrm{int},i},
\label{eq:absmagstar}
\end{equation}
where the stretch $x_1$, colour $c$, and intrinsic scatter $\sigma_{\mathrm{int}}$ are randomly drawn from statistical distributions, and we use best-fit values of $\alpha = 0.14$, $\beta=3.1$ and $B$-band absolute magnitude $M_B = -19.05$ \citep{betoule2014}. We rescaled $M$ to match the Abacus fiducial cosmology using,
\begin{equation}
    M = -19.05 + 5 \log_{10} \left( \frac{h}{0.7} \right).
\end{equation}

We modelled the $x_1$ distribution using a bimodal Gaussian mixture considering the evolution of younger and older SN Ia progenitors as a function of redshift \citep{rigault2020,nicolas2021}. The $x_1$ distribution is given by,
\begin{equation}
\begin{split}
    & x_1(z)  = \delta(z) \times \mathcal{N}(\mu_1, \sigma^2_1) \, + \\ 
    & (1 - \delta(z)) \times 
    \left[ a \times \mathcal{N}(\mu_1, \sigma^2_1) + (1 - a) \times \mathcal{N}(\mu_2, \sigma^2_2) \right] ,
\end{split}
\end{equation}
where $a$ accounts for the relative effect of the younger and older SN Ia progenitors, and the evolving fraction of young SNe Ia is given by $\delta(z) = (K^{-1} \times (1+z)^{-2.8} + 1)^{-1}$, with coefficient $K = 0.87$ as outlined by \cite{rigault2020}.  We used parameters $\mu_1 = 0.37$, $\sigma_1 = 0.61$, $\mu_2 = -1.22$, $\sigma_2 = 0.56$ and $a = 0.51$ \citep{nicolas2021} to determine the stretch parameter for each SN Ia.

The colour parameter $c$ is modelled as an asymmetric Gaussian distribution from \citet{2017ApJ...836...56K} following the low redshift (G10) model, where $\bar{c} = -0.054$ is the maximum probability colour parameter, $\sigma_{-} = 0.043$ and $\sigma_{+} = 0.101$ are the asymmetric Gaussian widths of the distribution.  The colour distribution follows,
\begin{equation}
\begin{split}
    P(c) \propto \exp \left[ - \frac{(c - \bar{c})^2}{2 \sigma_{-}^2} \right], \, c \leq \bar{c} , \\
    P(c) \propto \exp \left[ - \frac{(c - \bar{c})^2}{2 \sigma_{+}^2} \right], \, c > \bar{c} .
\end{split}
\end{equation}
Finally, the intrinsic scatter $\sigma_{\mathrm{int}}$ is drawn from a Gaussian distribution with dispersion $\sigma_{M} = 0.12$.  After drawing the values of $x_1$, $c$ and $\sigma_{\mathrm{int}}$ for each SN Ia from the distributions described above, the absolute magnitude in Eq.~\ref{eq:absmagstar} is used to calculate the apparent magnitude in Eq.~\ref{eq:dm}. 

The Pantheon+ compilation consists of 18 individual sub-surveys, resulting in a complex angular and depth selection. For this work, we approximated the angular selection of Pantheon+ using a Healpix map of the distribution of the Pantheon+ dataset and using this map, sub-sampled the mock and random catalogues. Modelling the selection of each individual sub-survey would improve the robustness of this type of analysis and, although beyond the scope of this project, would be worth considering for potential future analysis.  However, to test the robustness of our results to the choice of magnitude threshold, we considered applying different magnitude cuts at $m < 16.0$, $m < 16.5$, and $m < 17$, which are representative of the depth of the different samples.  This test is hence a first step to consider whether selection effects impact our type of correlation measurement.

\section{Correlation function measurements}
\label{sec:measurements}

In this section we describe our estimators for the auto- and cross-correlations of magnitude fluctuations and galaxy positions, which may be linked to the underlying velocity correlation theory using the models outlined in Sec.~\ref{sec:theory}.

\subsection{Magnitude auto-correlation functions}

As discussed in Sec.~\ref{sec:veltomag}, magnitude fluctuations are linked to the underlying velocity fluctuations by the redshift-dependent scaling $\alpha(z)$ defined in Eq.~\ref{eq:veltomag}.  Predicting the magnitude fluctuation correlation function requires us to normalise the model velocity correlations by the average product $\langle \alpha_A \, \alpha_B \rangle$ between a pair of galaxies $A$ and $B$ within the separation range, which we also measure during our correlation estimation.  In Sec.~\ref{sec:validnoseleff} we will test the appropriate separation range to include when fitting the model to the data.

The magnitude correlation between two galaxies depends on the correlation between their radial velocity components.  Re-writing Eq.~\ref{eq:corrtensor} as the correlation for line-of-sight velocities of two galaxies, $u_A$ and $u_B$, leads to,
\begin{equation}
\begin{split}
    & \langle u_A(\mathbf{r}_A) \, u_B(\mathbf{r}_B) \rangle = \\ & \Psi_{\perp}(r) \cos\theta_{AB} + \left[ \Psi_{\parallel}(r) - \Psi_{\perp}(r) \right] \cos\theta_A \cos\theta_B ,
\end{split}
\label{eq:corrradial}
\end{equation}
where $\theta_A$ and $\theta_B$ are the angles between the lines-of-sight to the two objects and the separation vector, and $\theta_{AB}$ is the angle subtended by the two lines-of-sight at the origin.  Converting Eq.~\ref{eq:corrradial} to the corresponding magnitude fluctuations $\delta m_A$ and $\delta m_B$ we find,
\begin{equation}
\label{eq:corrmag}
    \langle \delta m_A(\mathbf{r}_A) \, \delta m_B(\mathbf{r}_B) \rangle = \alpha_A \, \alpha_B \langle u_A(\mathbf{r}_A) \, u_B(\mathbf{r}_B) \rangle .
\end{equation}

The dependence of Eq.~\ref{eq:corrradial} on the orientation of the galaxy pair with respect to the line-of-sight shows that the auto-correlation spectrum must be characterised by two separate functions \citep{2024MNRAS.527..501B}.  \citet{gorski1989} demonstrated that the information in Eq.~\ref{eq:corrradial} can be written in terms of the two correlation statistics $\psi_1$ and $\psi_2$.  Using magnitude fluctuations as the variable, these may be estimated as,
\begin{align}
\begin{split} \label{eq:psi1}
    \hat{\psi}_1(r) ={}& \frac{\sum w_A \, w_B \, \delta m_A \, \delta m_B \, \cos\theta_{AB}}{\sum w_A \, w_B \, \cos^2\theta_{AB}} ,
\end{split} \\
\begin{split} \label{eq:psi2}
    \hat{\psi}_2(r) ={}& \frac{\sum w_A \, w_B \, \delta m_A \, \delta m_B \, \cos\theta_A \, \cos\theta_B}{\sum w_A \, w_B \, \cos\theta_{AB} \, \cos\theta_A \, \cos\theta_B} ,
\end{split}
\end{align}
where sums are over all pairs of galaxies separated by some fixed distance $r$, and galaxies are weighted by $w_A$ and $w_B$, discussed further below. These estimators for $\psi_1$ and $\psi_2$ can be related to the velocity correlation functions $\Psi_{\parallel}$ and $\Psi_{\perp}$ using Eqs.~\ref{eq:corrradial}, \ref{eq:corrmag}, \ref{eq:psi1} and \ref{eq:psi2} to give,
\begin{equation}
    \langle \psi_1(r) \rangle = \frac{N^{\mathrm{auto}}_{\alpha\alpha}}{N^{\mathrm{auto}}_{gg}} \, \left\{ \mathcal{A}(r) \psi_{\parallel}(r) + \left[ 1 - \mathcal{A}(r) \right] \psi_{\perp}(r) \right\},
\end{equation}
and,
\begin{equation}
    \langle \psi_2(r) \rangle = \frac{N^{\mathrm{auto}}_{\alpha\alpha}}{N^{\mathrm{auto}}_{gg}} \, \left\{ \mathcal{B}(r) \psi_{\parallel}(r) + \left[ 1 - \mathcal{B}(r) \right] \psi_{\perp}(r) \right\},
\end{equation}
where $\mathcal{A}(r)$ and $\mathcal{B}(r)$ are geometry factors dictating the relative contributions of the parallel and perpendicular components of the velocity field to the correlation functions $\psi_1$ and $\psi_2$, which are given by \cite{gorski1988},
\begin{equation}
    \mathcal{A}(r) = \frac{\sum w_A \, w_B \, \cos \theta_A \, \cos \theta_B \, \cos \theta_{AB}}{\sum w_A \, w_B \cos^2 \theta_{AB}},
\end{equation}
\begin{equation}
    \mathcal{B}(r) = \frac{\sum w_A \, w_B \, \cos \theta_A \, \cos \theta_B \, \cos \theta_{AB}}{\sum w_A \, w_B \, \cos^2\theta_A}.
\end{equation}
The normalisation factor $N^{\mathrm{auto}}_{gg}$ is evaluated by taking the velocity-velocity pair count in each separation bin weighted by the velocity sample weight $w_v$ for each galaxy in the pair. The factor $N^{\mathrm{auto}}_{\alpha\alpha}$ is evaluated by taking the velocity-velocity pair count weighted by $w_v$ and the values of $\alpha(z)$ for each galaxy.

The optimal weights for each galaxy are chosen to minimize the statistical error in the 2-point correlation function, balancing sample variance and measurement noise \citep{1994ApJ...426...23F}, given the varying number density of galaxies across the survey volume. The optimal weight for the galaxy sample is,
\begin{equation}
    w_{g} = \frac{1}{1 + n_g \, P_g},
\end{equation}
where $n_g$ is the number density in units of \unit{\h^3.Mpc^{-3}} of the galaxy sample at the position in question, and the characteristic galaxy power spectrum amplitude is taken as $P_g = 10^4$~\unit{\h^{-3}.Mpc^3}.
The optimal weight for the velocity sample is \citep[e.g.,][]{2023MNRAS.518.2436T},
\begin{equation}
    w_{v} = \frac{1}{\sigma^2_m / \alpha(z) + \alpha(z) \, n_v \, P_v},
\end{equation}
where $n_v$ is the number density of the velocity sample in \unit{\h^3 Mpc^{-3}}, the characteristic velocity power spectrum amplitude is taken as $P_v = 10^{10}$~\unit{\h^{-3}.Mpc^3.(km.s^{-1})^2}, $\sigma_m$ is the supernova-specific magnitude error, and $\alpha(z)$ is the redshift-dependent scaling factor defined in Eq.~\ref{eq:veltomag}, evaluated for each galaxy.

\begin{figure*}
    \centering
    \includegraphics[width=\textwidth]{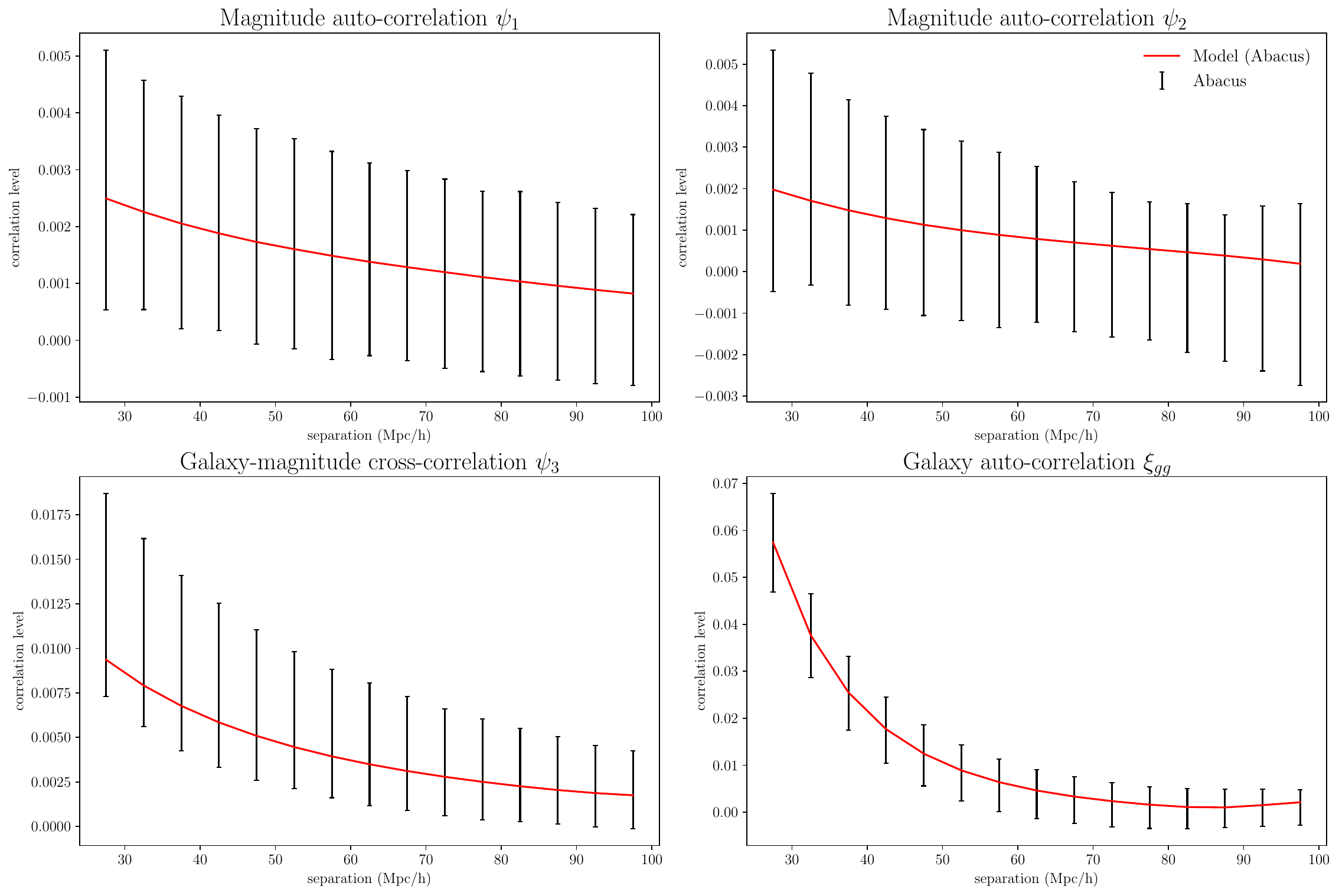}
    \caption{The $\left( \psi_1, \psi_2, \psi_3, \xi_{gg} \right)$ correlation functions in our tests using the Abacus simulations (Sec.~\ref{sec:abacus}).  We show measurements of the mock mean and standard deviation over the realisations as the data points, and best-fit models as the solid lines.}
    \label{fig:mockcorr}
\end{figure*}

\begin{figure*}
    \centering
    \includegraphics[width=\textwidth]{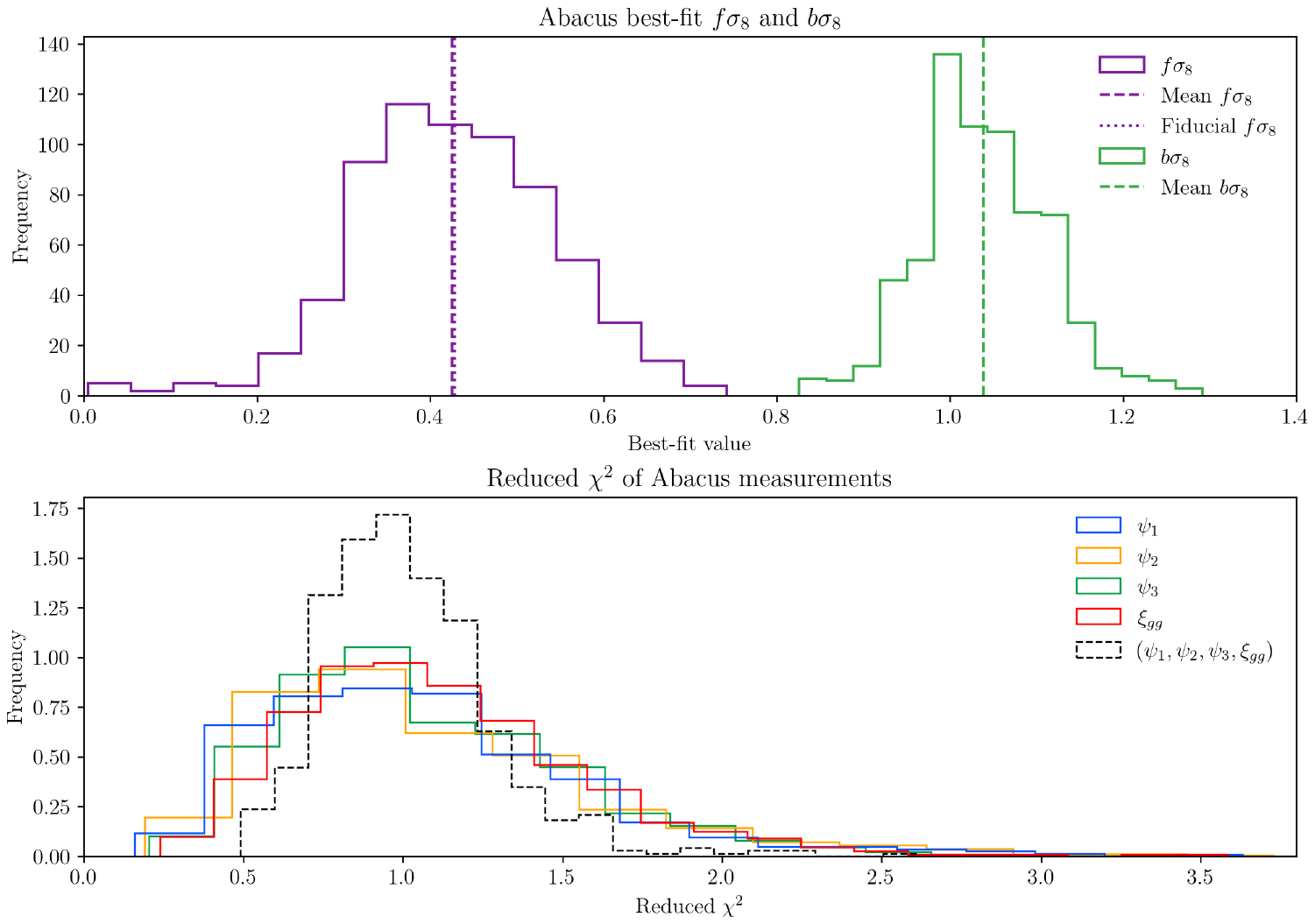}
    \caption{Mock validation results, using the 675 Abacus mocks without SN selection effects. Top: histogram of the best-fit $f \sigma_8$ and $b \sigma_8$ values from the combined four-statistic fit. Bottom: histogram of the reduced $\chi^2$ values for fits to the individual- and combined four- correlation function measurements.}
    \label{fig:mockfits}
\end{figure*}

\begin{figure}
    \centering
    \includegraphics[width=\columnwidth]{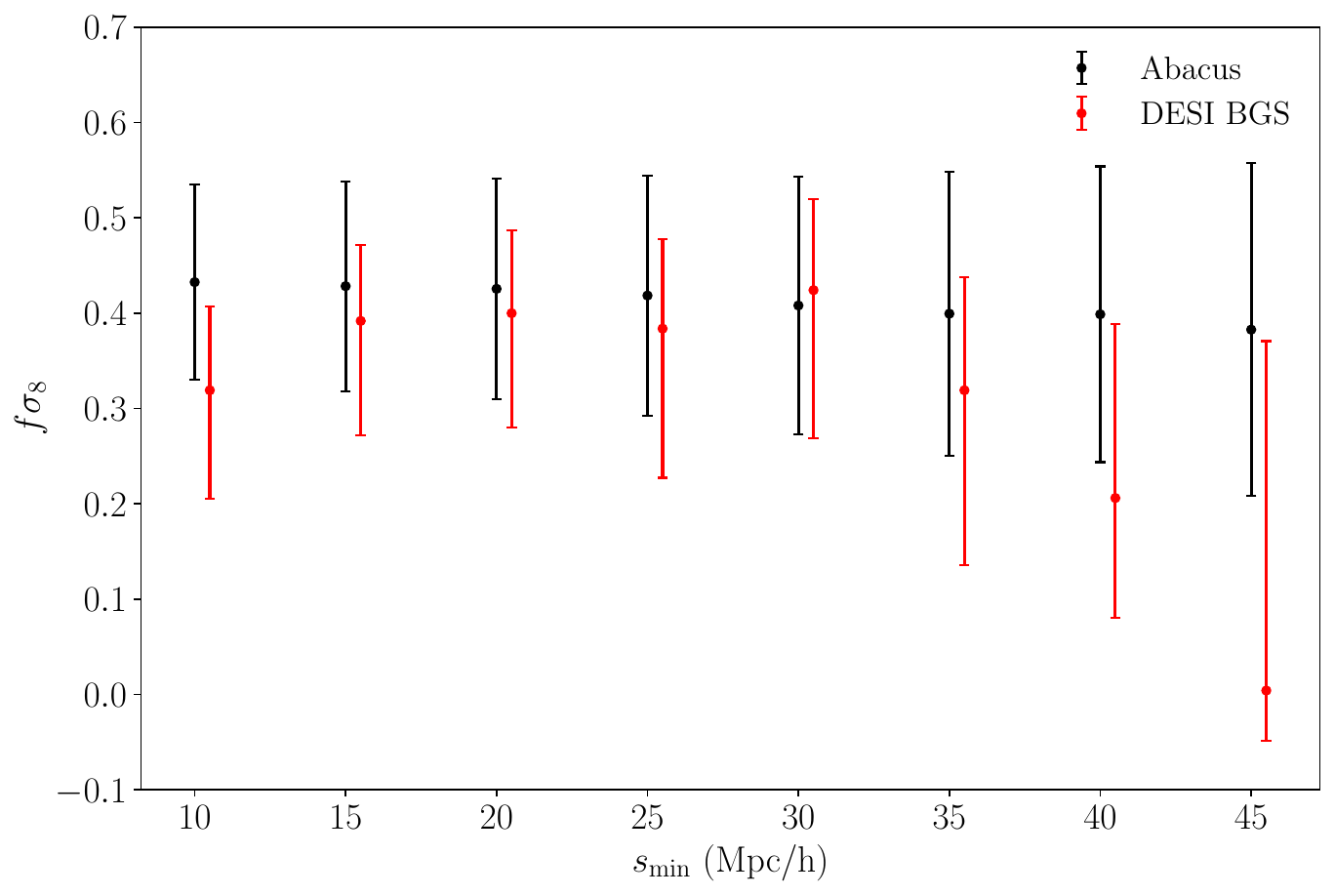}
    \caption{The best-fit $f \sigma_8$ and $68\%$ confidence region for the Abacus mock (black points, mean and standard deviation across 675 realisations) and the DESI BGS and Pantheon+ data (red points, best-fit and error) for a range of minimum fitted separations ($10 - 45 \, h^{-1}$\unit{Mpc}).  The maximum fitted separation is always $100 \, h^{-1}$\unit{Mpc}.}
    \label{fig:scaletest}
\end{figure}

\begin{figure*}
    \centering
    \includegraphics[width=\textwidth]{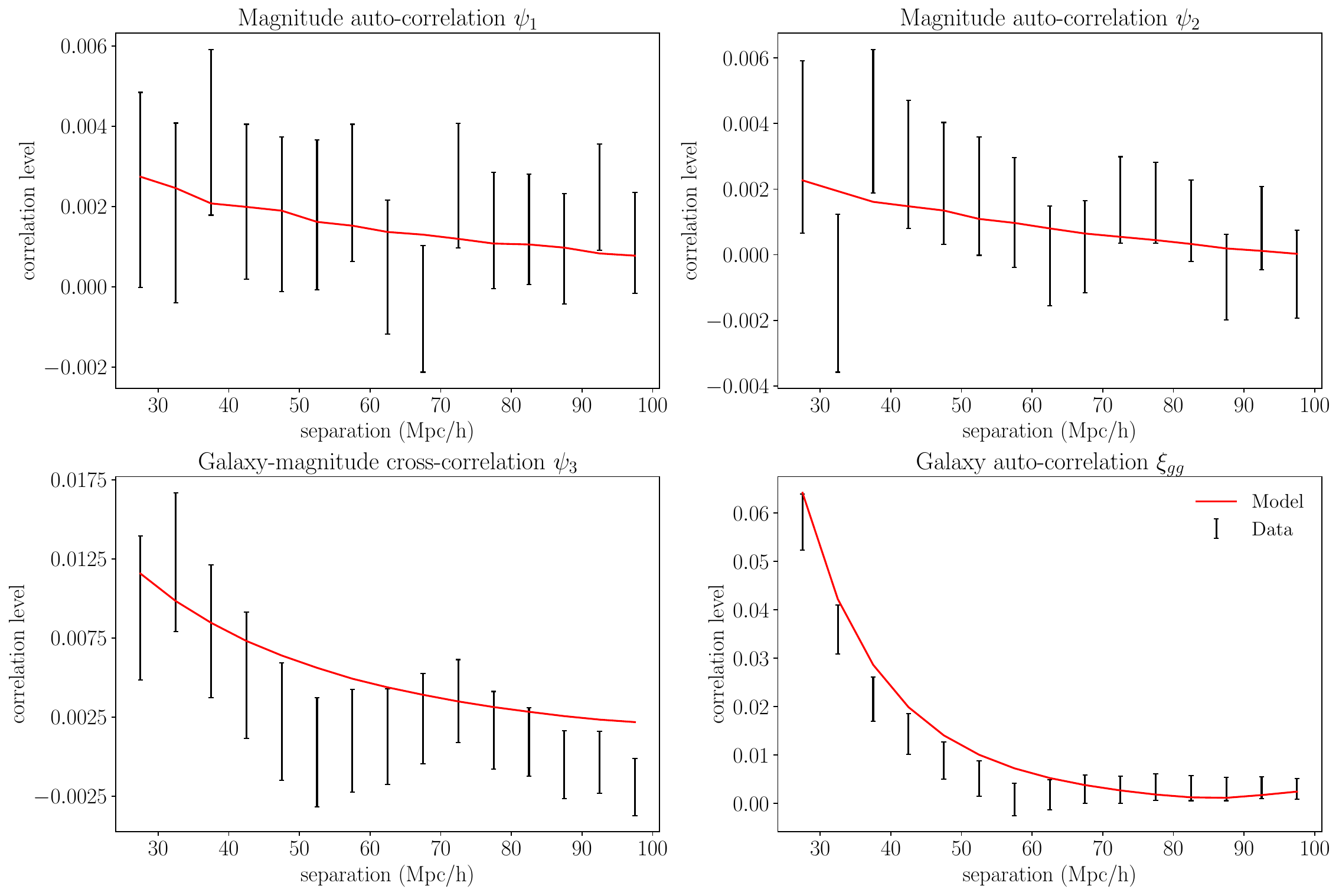}
    \caption{The $\left( \psi_1, \psi_2, \psi_3, \xi_{gg} \right)$ correlation functions for the DESI and Pantheon+ samples, displayed in different panels and overplotted with the best-fit models.  The error bars are shown as the standard deviation of the mock measurements, and the best-fit models are shown as the red lines.}
    \label{fig:datacorr}
\end{figure*}

\subsection{Galaxy-magnitude cross-correlation function}

Converting Eq.~\ref{eq:galvelcorr} to a line-of-sight velocity, the cross-correlation function between a galaxy overdensity at position $A$ and magnitude fluctuation at position $B$ is,
\begin{equation}
    \langle \delta_A (\mathbf{r}_A) \, \delta m_B (\mathbf{r}_B) \rangle = \alpha_B \, \xi_{gv}(r) \cos\theta_B.
\end{equation}
The estimator for the cross-correlation function between the galaxy overdensity and magnitude fluctuations can then be defined following \citet{turner2021} as,
\begin{equation}
    \hat{\psi_3}(r) = \frac{\sum w_A \, w_B \, \delta m_B \, \cos\theta_B}{\sum w_A \, w_B \, \cos^2\theta_B} .
\label{eq:psi3}
\end{equation}
We form the complete cross-correlation estimator by subtracting from Eq.~\ref{eq:psi3} the analogous cross-pair count between the velocity sample and a distribution of unclustered random galaxies with the same distribution as the galaxy data.  In this way, the variance of the galaxy-magnitude cross-correlation estimator is significantly reduced \citep{turner2021}.

To relate the theoretical galaxy-velocity cross-correlation model from Eq.~\ref{eq:galvelcorr} to the galaxy-magnitude cross-correlation function measurements, we normalise the model as,
\begin{equation}
    \langle \hat{\psi_3}(r)\rangle = \frac{N^{\mathrm{cross}}_{g\alpha}}{N^{\mathrm{cross}}_{gg}} \, \xi_{gv}(r) .
\end{equation}
The normalisation factor $N^{\mathrm{cross}}_{gg}$ is evaluated by taking the cross-pair count of the galaxy and velocity samples in each separation bin weighted by the galaxy and velocity optimal weights, $w_g$ and $w_v$, respectively.  The factor $N^{\mathrm{cross}}_{g \alpha}$ is evaluated by taking the galaxy-velocity pair count weighted by the optimal weights of each pair and an additional factor of $\alpha(z)$.

\subsection{Galaxy auto-correlation function}

The galaxy auto-correlation function estimator is given by,
\begin{equation}
    \hat{\xi}_{gg}(r) = \frac{N^2_R}{N^2_D} \frac{D_g D_g(r)}{R_g R_g(r)} - 2 \frac{N_R}{N_D} \frac{D_g R_g(r)}{R_g R_g(r)} + 1,
\label{eq:xigg}
\end{equation}
\citep{landy1993}, where $D_g D_g(r)$ and $R_g R_g(r)$ are the galaxy-galaxy pair count at separation $r$, for the data and random samples, respectively, and $D_g R_g(r)$ is the cross-pair count between the samples. $N_D$ and $N_R$ are the (weighted) number of galaxies in the data and random sample, respectively and are used here to normalise the pair count.  The inclusion of the second term in Eq.~\ref{eq:xigg} decreases the statistical error associated with the distribution of the data with respect to sample boundaries \citep{landy1993}.

\section{Validation using simulations}
\label{sec:validation}

In this section we present the suite of tests performed using N-body simulations to validate the recovery of the fiducial growth rate using our correlation measurements and models.

\subsection{Validation by fitting to Abacus simulations (without SN selection effects)}
\label{sec:validnoseleff}

As described in Sec.~\ref{sec:abacus}, the Abacus mock catalogues were constructed by applying a selection function matching the DESI BGS and a maximum redshift of $z = 0.1$. The SN mock catalogues were matched to the Pantheon+ dataset by sub-sampling the parent catalogues by redshift and sky position, and an intrinsic dispersion $\sigma_{\mathrm{int}} = 0.126$ was added to the SN magnitudes, which (after the inclusion of peculiar velocities) matches the standard deviation of the observed Pantheon+ magnitudes around the fiducial magnitude-redshift relation.

For our first test, we excluded supernovae selection effects.  For each of the 675 mock catalogues we measured the correlation functions $\left( \psi_1, \psi_2, \psi_3, \xi_{gg} \right)$ in a separation range of $25 - 100$~\unit{\h^{-1}.Mpc} in 5~\unit{\h^{-1}.Mpc} bins, where the minimum separation is an estimate of the range of applicability of linear theory (which we test further below), and the maximum separation marks a reduction in the number of available SN pairs.  In Fig.~\ref{fig:mockcorr} we display these statistics averaged over the catalogues.  We also used this ensemble of measurements to construct a numerical covariance matrix of the statistics.

To find the best-fitting values of the normalised growth rate $f \sigma_8$ and galaxy bias $b \sigma_8$, we fit the correlation function measurements using the likelihood function $\mathcal{L} \propto \exp(-\chi^2/2)$ in terms of the $\chi^2$ function,
\begin{equation}
    \chi^2 = \sum_i \sum_j (d_i - m_i)^T \, \left( C^{-1} \right)_{ij} \, (d_j - m_j) ,
\label{eq:chi2}
\end{equation}
where $d_i$ and $m_i$ are concatenations of the four correlation function datasets and models, respectively, and $C$ is the covariance matrix, and minimising the $\chi^2$ statistic for each realisation. 

The histogram of the best-fit $f \sigma_8$ and $b \sigma_8$ values across the realisations is displayed in the top panel of Fig.~\ref{fig:mockfits}, and the mean of the marginalised posterior distribution for $f \sigma_8$ with the $\chi^2$ and degrees of freedom are shown in Table \ref{tab:correlationmock} for different combinations of correlation functions. The distribution of reduced $\chi^2$ values for the fits to each of the correlation functions is shown in the bottom panel of Fig.~\ref{fig:mockfits}.  All correlations result in acceptable $\chi^2$ statistics, validating the sufficiency of our covariance and modelling approach. The growth rate fit for the combined correlation function measurements gives $f \sigma_8 = 0.425^{+0.089}_{-0.150}$ where we quote the mean and standard deviation across 675 realisations. The equivalent statistics for the galaxy bias parameter are $b \sigma_8 = 1.042^{+0.065}_{-0.084}$. Agreement between the measurements and model across the four correlation functions demonstrates that our methodology is capable of recovering the fiducial value $f \sigma_8 = 0.428$ within the statistical error margin, in the presence of realistic survey selection functions, galaxy bias and measurement noise.

In our fiducial analysis we restricted the fitting range of the correlation functions to $25-100$~\unit{\h^{-1}.Mpc} to mitigate the impact of non-linear growth violating our linear modelling assumptions. To determine the impact of this scale cut on the growth rate measurements, we also fit to the combined four correlation functions of the mocks varying the minimum separation in the range $10-45$~\unit{\h^{-1}.Mpc}, finding that the best-fit $f \sigma_8$ values were consistent across different separation ranges. Although there appears to be a weak trend for decreasing $f\sigma_8$ as $s_{\mathrm{min}}$ increases for measurements using the real data in Fig.~\ref{fig:scaletest}, the variation in results between each separation range is significantly smaller than the statistical uncertainty for the analysis, such that the choice of minimum separation bin leads to a statistically indistinguishable result. This demonstrates that our growth rate fits are robust to the choice of fitting range. At the maximum fitting range of 100~\unit{\h^{-1}.Mpc}, the SN number density has diminished (Fig.~\ref{fig:reddist}) and the low quantity of SN pairs limits the potential signal-to-noise at larger scales.

\begin{table}[ht]
    \centering
    {\renewcommand{\arraystretch}{1.73}%
    \begin{tabular}{c|c|c|c}
    Correlation function & $f \sigma_8$ & $\chi^2$ & d.o.f. \\
    \hline
    $\psi_1$ & $0.422^{+0.144}_{-0.220}$ & 11.35 & 13 \\
    $\psi_2$ & $0.352^{+0.218}_{-0.183}$ & 11.57 & 13 \\
    $\psi_3 + \xi_{gg}$ & $0.538^{+0.513}_{-0.280}$ & 23.06 & 28 \\
    combined & $0.425^{+0.089}_{-0.150}$ & 47.65 & 58
    \end{tabular}}
    \caption{The $68\%$ confidence intervals of the $f \sigma_8$ marginalised posterior distribution, minimum $\chi^2$ and number of degrees of freedom (d.o.f.) when fitting to different combinations of correlation functions using the Abacus mocks, averaged over 675 mocks.}
    \label{tab:correlationmock}
\end{table}

\subsection{Validation including SN selection effect}

To test the impact of SN selection effects on magnitude correlations, we used SN catalogues with and without selection effects applied, as described in Sec.~\ref{sec:seleff}.  We apply magnitude limits of $m = (16.0, 16.5, 17.0)$ and fit our correlation function models to the 675 Abacus realisations.  In each case, we measured the mean difference in the best-fitting $f\sigma_8$ with and without selection effects applied.  For the three magnitude thresholds $m = (16.0, 16.5, 17.0)$ we found a mean difference in $f \sigma_8$ of $(0.027, 0.016, 0.0064)$.  These offsets are small compared to the statistical errors, demonstrating that SN selection effects do not significantly bias our growth rate measurements at the level of precision of these datasets.


\section{Growth rate fits to DESI and Pantheon+}
\label{sec:fits}

Having validated our analysis pipeline using mock catalogues, we now apply it to the DESI DR1 and Pantheon+ datasets.  In the redshift range $z < 0.1$, our sample consists of 578,576 DESI BGS galaxies and 510 Pantheon+ SNe Ia.  We measured the auto- and cross-correlation statistics between the two samples in the separation range $25 - 100$~\unit{\h^{-1}.Mpc} in 5~\unit{\h^{-1}.Mpc} bins. These correlation measurements are shown in Fig.~\ref{fig:datacorr}.

\begin{figure}
\centering
\includegraphics[width=\columnwidth]{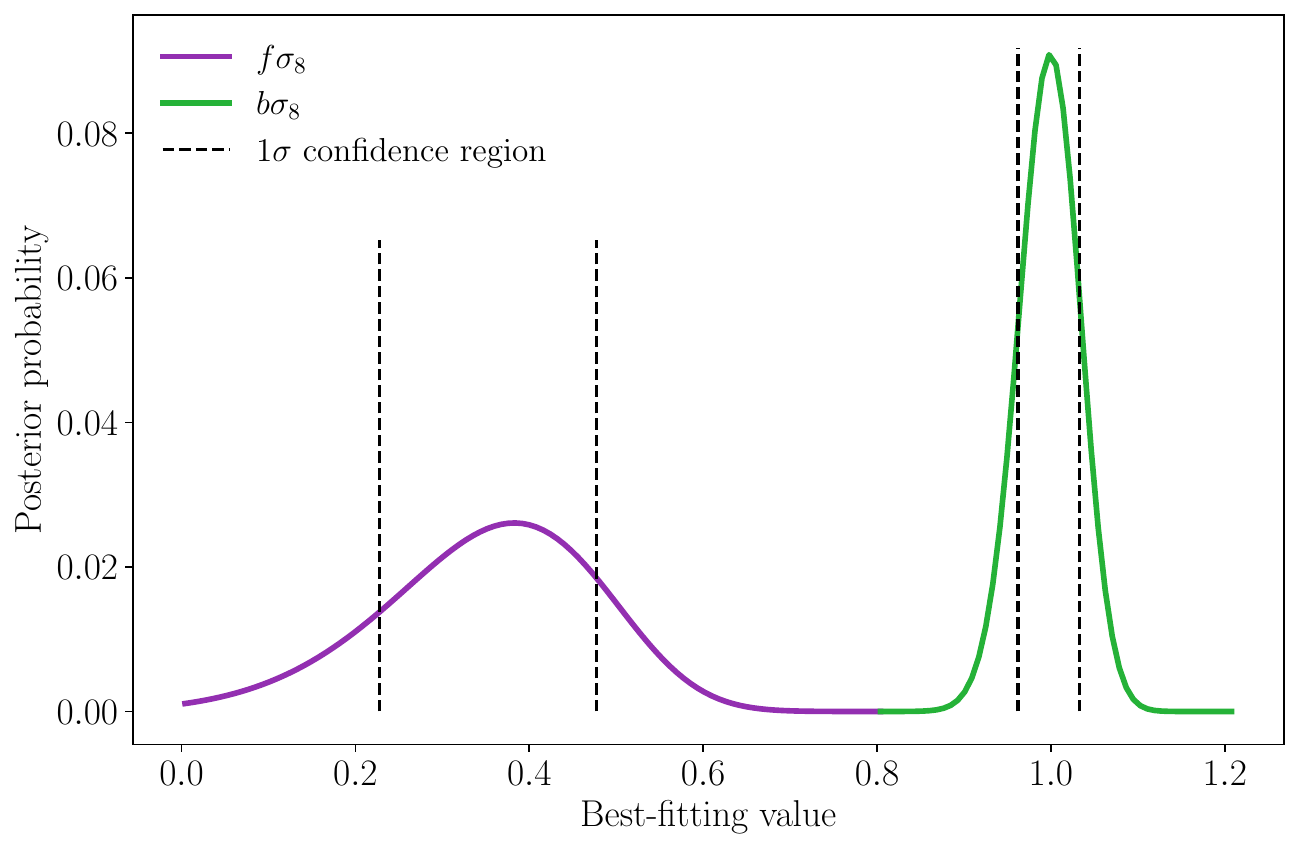}
\caption{The posterior probability distributions for the normalised growth rate $f \sigma_8$ and normalised galaxy bias $b \sigma_8$ fit to the two-point correlation function measurements between DESI BGS galaxies and Pantheon+ magnitude fluctuations.}
\label{fig:datafits}
\end{figure}

We fit our correlation models varying $f \sigma_8$ and $b \sigma_8$ as before, evaluating $\chi^2$ using Eq.~\ref{eq:chi2}.  We show the marginalised posterior probability distributions for $f \sigma_8$ and $b \sigma_8$ from the combined four-statistic fit to the data catalogues in Fig.~\ref{fig:datafits}. The $68\%$ confidence interval $f \sigma_8 = 0.384^{+0.094}_{-0.157}$ at $z = 0.03$ agrees with the prediction of the {\it Planck} cosmology $f \sigma_8 = 0.429$, and the $68\%$ confidence interval for the bias parameter is $b \sigma_8 = 0.998^{+0.035}_{-0.036}$.  For the combined $f \sigma_8$ and $b \sigma_8$ constraint, the minimum $\chi^2 = 62.59$, consistent with the expected value from a fit with 58 degrees of freedom.  We note that full-shape galaxy clustering using the full DESI BGS sample measured $f \sigma_8 = 0.38 \pm 0.09$ \citep{2025JCAP...09..008A} using redshift-space distortion information, and our constraints, which do not use redshift-space distortion information in the galaxy correlation function, are consistent with this measurement.  In Table \ref{tab:correlationdata} we list the $68\%$ confidence interval of the $f \sigma_8$ marginalised posterior distribution and minimum $\chi^2$ values for different combinations of correlation functions, demonstrating that the models provide a good fit to all correlations.  The effect of the fitting range on the $f\sigma_8$ measurements is shown for the DESI+Pantheon correlations in Fig.~\ref{fig:scaletest}, which demonstrates that the minimum separation does not significantly affect the $f\sigma_8$ values.

\begin{figure*}
\centering
\includegraphics[width=\textwidth]{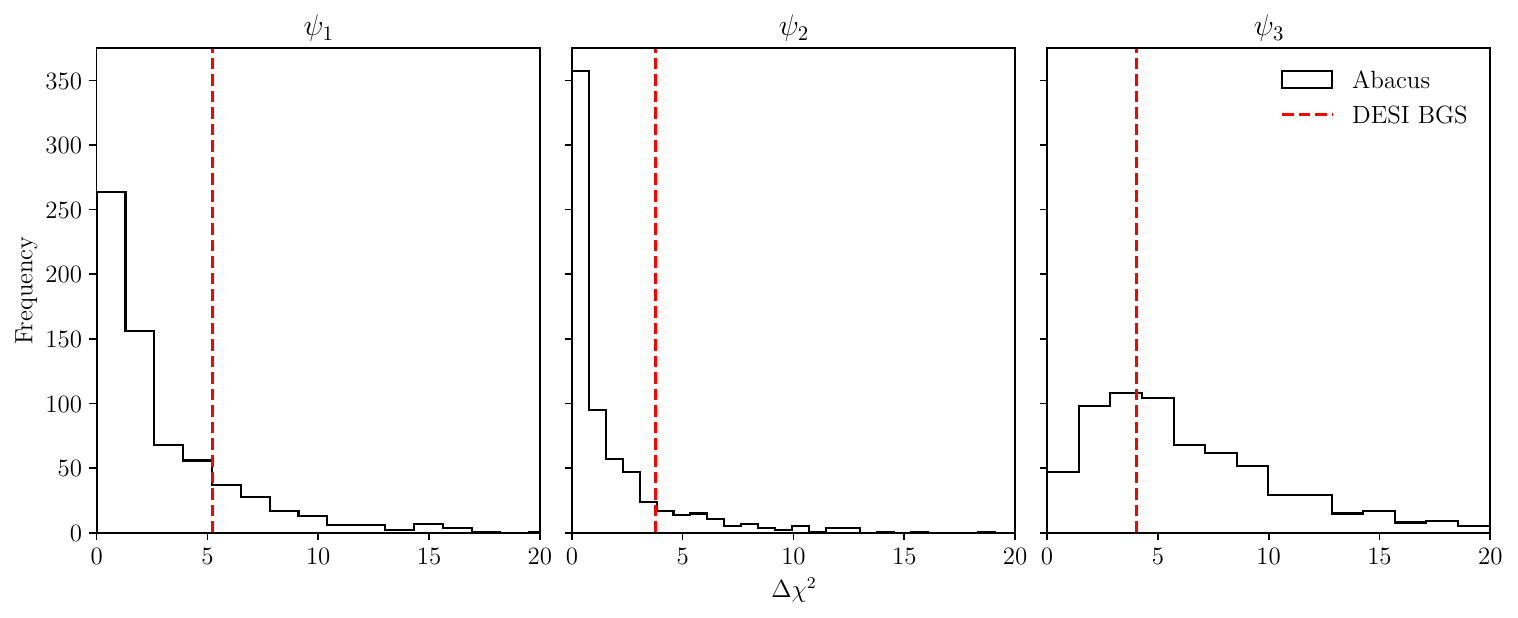}
\caption{The $\Delta \chi^2$ difference between the best-fitting model and a zero-correlation model, considered separately for each correlation function, for the ensemble of Abacus mocks (represented by the black histogram) and for the DESI BGS and Pantheon+ datasets (represented by the vertical red dashed line).  This difference quantifies the extent to which magnitude correlations are ``detected'' in the samples.}
\label{fig:deltachisq}
\end{figure*}

As a measure of the extent to which we ``detect'' the magnitude correlations in our samples, we considered the $\Delta \chi^2$ between the best-fitting $f \sigma_8$ model and a zero-correlation model separately for the $\psi_1$, $\psi_2$ and $\psi_3$ functions, for the ensemble of Abacus mocks and the DESI BGS and Pantheon+ datasets.  The results are shown in Fig.~\ref{fig:deltachisq}, with the distribution of mock results displayed as the histograms, and the DESI-Pantheon+ measurement shown as the vertical lines.  The differences $\Delta \chi^2 \sim 4$ for the DESI-Pantheon+ measurements represent a $\sim 2$-$\sigma$ detection for each individual correlation, which accumulates across the different statistics.  The results for the ensemble of mocks show that a wide range of detection significances may be obtained.  Our results for the real data are consistent with these distributions, resulting in a somewhat more significant detection than the mock average for the auto-correlations $\psi_1$ and $\psi_2$, and a somewhat less significant detection for the cross-correlation $\psi_3$.

This is the first detection of the correlation between the fluctuations in SNe magnitudes and large-scale structure.  Our results are consistent with {\it Planck} predictions, but due to the large uncertainties we cannot distinguish between $\Lambda$CDM and the dynamical dark energy model $w_0 w_a$CDM favoured by \citet{2025arXiv250314738D}.

\begin{table}[ht]
    \centering
    {\renewcommand{\arraystretch}{1.73}%
    \begin{tabular}{c|c|c|c}
    Correlation function & $f \sigma_8$ & $\chi^2$ & d.o.f. \\
    \hline
    $\psi_1$ & $0.400^{+0.092}_{-0.242}$ & 12.06 & 13 \\
    $\psi_2$ & $0.440^{+0.093}_{-0.270}$ & 12.44 & 13 \\
    $\psi_3 + \xi_{gg}$ & $0.315^{+0.299}_{-0.146}$ & 36.41 & 28 \\
    combined & $0.384^{+0.094}_{-0.157}$ & 62.59 & 58 \\
    \end{tabular}}
    \caption{The $68\%$ confidence intervals of the $f \sigma_8$ marginalised posterior distribution, minimum $\chi^2$ and number of degrees of freedom for our model fits to different combinations of correlation functions using the DESI BGS and Pantheon+ datasets.}
    \label{tab:correlationdata}
\end{table}

\subsection{Forecasting improvements in $f\sigma_8$ constraints with future SN samples}
\label{sec:forecast}

The statistical significance of these correlations, and the accuracy with which they can constrain the growth rate, will improve as galaxy and SN samples grow in size. To forecast these improvements and demonstrate the potential of this method, we performed Fisher matrix forecasts of the growth rate error for different SN sample sizes, otherwise maintaining the properties of the DESI BGS and Pantheon+ configurations.  We consider the combined $(\psi_1, \psi_2, \psi_3)$ statistics varying the growth rate $f$ and evaluate the Fisher matrix value as,
\begin{equation}
\label{eq:fismat}
    F = \frac{\partial \mathbf{m}^T}{\partial f}  \mathbf{C}^{-1} \frac{\partial \mathbf{m}}{\partial f},
\end{equation}
where $\mathbf{m}$ is the correlation function model and $\mathbf{C}$ is the covariance matrix.  For the purpose of this test, we used Gaussian analytical covariances using the expressions in \cite{2024MNRAS.527..501B} which include the effects of sample variance, selection functions of the density and velocity tracers, shot noise and velocity measurement errors and curved-sky effects.  \cite{2024MNRAS.527..501B} demonstrate that these expressions are accurate representations of the covariance across statistics and scales, if the selection function of the datasets does not significantly vary on the separation scale in question.  We considered the $(\psi_1, \psi_2, \psi_3)$ statistics varying the growth rate $f$, where the error forecast for $f$ is given by $F^{-1/2}$.  We considered SN samples of size $( 100, 300, 1000, 3000 )$, finding the forecast fractional errors in $f \sigma_8$ to be $( 49.8\%, 22.5\%, 7.7\%, 3.0\% )$ for these cases.

We note that the forecast error is somewhat more optimistic than the result of our analysis of the real data.  Two important reasons for this discrepancy are, first, that the forecast is based on an analytical covariance, and the data measurement uses a mock covariance, where the latter should be more reliable.  Secondly, a Fisher matrix forecast provides the minimum possible error, and this could highlight that the weighting used in the analysis could be further optimised.  Whilst a simplified forecast, the results lie within a factor of 2 of the growth rate measurement using the real sample, and the relative scaling with SNe number should be robust.  This analysis demonstrates that future SN surveys have the potential to improve growth rate errors by a further order of magnitude (for more detailed forecasts, see \cite{howlett2017} and \cite{2023A&A...674A.197C}).

Given the signal-to-noise offered by the current SN samples, we have used a simplified linear model in this study. However, given future SN surveys with increased signal-to-noise, a fitting model including parameters for non-linear corrections \citep[e.g.][]{2025arXiv251203231Q}, including non-linear galaxy bias and velocity dispersions, will allow for more accurate tests of cosmology.  Given more accurate statistical errors, the impact of SN selection effects will also become relatively more important to model precisely.

\section{Conclusions}
\label{sec:disc}

In this study we report a modest detection of the magnitude correlations within current SN samples, and between SNe and local large-scale structure, using the DESI Bright Galaxy Survey and Pantheon+ datasets.  These correlations are associated with peculiar velocities induced by local large-scale structure, and may be used to measure the growth rate of structure.  The $68\%$ confidence intervals of the marginalised posterior distribution from the correlation fit to the data catalogues are $f \sigma_8 = 0.384^{+0.094}_{-0.157}$ and $b \sigma_8 =0.998^{+0.035}_{-0.036}$ at $z = 0.03$. This measured growth rate value is consistent with expectations of the $\Lambda$CDM cosmological model.  At the 1-$\sigma$ level, our measurement also agrees with the DESI full-shape redshift-space distortion analysis of the BGS sample, which finds $f \sigma_8 = 0.38 \pm 0.09$ \citep{2025JCAP...09..008A}.

We used realistic mock catalogues, drawn from the Abacus simulations, to validate our methodology and demonstrate that we are able to recover the fiducial growth rate within the statistical errors in the presence of realistic survey selection functions, galaxy bias and measurement noise.  We also demonstrated that the SN selection effects have a negligible impact on the growth rate measurements at the level of precision of this analysis.

The forthcoming combination of the DESI BGS dataset and DESI Peculiar Velocity survey, which contains both Fundamental Plane and Tully-Fisher samples \citep{2023MNRAS.525.1106S, 2025MNRAS.539.3627S, 2025arXiv250711765D}, will enable accurate tests of the growth of structure.  The next generation of transient surveys such as the Zwicky Transient Facility \citep{2019PASP..131a8002B, 2025A&A...694A...1R} and Rubin Observatory's Legacy Survey of Space and Time \citep{2019ApJ...873..111I, Rosselli_LSST_2025} will provide a 1-2 order-of-magnitude increase in the number of SNe detected, allowing us to achieve unprecedented precision in testing cosmological models using supernovae \citep{howlett2017, 2020PDU....2900519G, 2023A&A...674A.197C}.  This expansion in sample size will necessitate extensions of multiple aspects of our analysis including non-linear modelling of the power spectrum, covariance modelling and the impact of SN selection effects.

\section*{Acknowledgements}

We thank Anais M\"{o}ller for useful discussions of supernovae selection effects, and Greg Aldering for helpful comments on a draft of this paper.

This research was conducted by the Australian Research Council Centre of Excellence for Gravitational Wave Discovery (project number CE230100016) and funded by the Australian Government.  AN would like to acknowledge the financial support received through the award of a Research Training Program Stipend scholarship by Swinburne University.

The project leading to this publication has received funding from 
Excellence Initiative of Aix-Marseille University - A*MIDEX, 
a French ``Investissements d'Avenir'' program (AMX-20-CE-02 - DARKUNI).

This material is based upon work supported by the U.S. Department of Energy (DOE), Office of Science, Office of High-Energy Physics, under Contract No. DE–AC02–05CH11231, and by the National Energy Research Scientific Computing Center, a DOE Office of Science User Facility under the same contract. Additional support for DESI was provided by the U.S. National Science Foundation (NSF), Division of Astronomical Sciences under Contract No. AST-0950945 to the NSF’s National Optical-Infrared Astronomy Research Laboratory; the Science and Technology Facilities Council of the United Kingdom; the Gordon and Betty Moore Foundation; the Heising-Simons Foundation; the French Alternative Energies and Atomic Energy Commission (CEA); the National Council of Humanities, Science and Technology of Mexico (CONAHCYT); the Ministry of Science, Innovation and Universities of Spain (MICIU/AEI/10.13039/501100011033), and by the DESI Member Institutions: \url{https://www.desi.lbl.gov/collaborating-institutions}. Any opinions, findings, and conclusions or recommendations expressed in this material are those of the author(s) and do not necessarily reflect the views of the U. S. National Science Foundation, the U. S. Department of Energy, or any of the listed funding agencies.

The authors are honored to be permitted to conduct scientific research on I'oligam Du'ag (Kitt Peak), a mountain with particular significance to the Tohono O’odham Nation.

\section*{Data availability}

Data points for all the figures are available at \url{https://doi.org/10.5281/zenodo.17111172}.

\bibliographystyle{mnras}
\bibliography{main}

@ARTICLE{2023MNRAS.518.2436T,
       author = {{Turner}, Ryan J. and {Blake}, Chris and {Ruggeri}, Rossana},
        title = "{A local measurement of the growth rate from peculiar velocities and galaxy clustering correlations in the 6dF Galaxy Survey}",
      journal = {\mnras},
         year = 2023,
        month = jan,
       volume = {518},
       number = {2},
        pages = {2436-2452},
          doi = {10.1093/mnras/stac3256},
archivePrefix = {arXiv},
       eprint = {2207.03707},
 primaryClass = {astro-ph.CO},
       adsurl = {https://ui.adsabs.harvard.edu/abs/2023MNRAS.518.2436T}
}

@ARTICLE{hui2006,
       author = {{Hui}, Lam and {Greene}, Patrick B.},
        title = "{Correlated fluctuations in luminosity distance and the importance of peculiar motion in supernova surveys}",
      journal = {\prd},
         year = 2006,
        month = jun,
       volume = {73},
       number = {12},
          eid = {123526},
        pages = {123526},
          doi = {10.1103/PhysRevD.73.123526},
archivePrefix = {arXiv},
       eprint = {astro-ph/0512159},
 primaryClass = {astro-ph},
       adsurl = {https://ui.adsabs.harvard.edu/abs/2006PhRvD..73l3526H}
}

@ARTICLE{2019MNRAS.490.2948D,
       author = {{Davis}, Tamara M. and {Hinton}, Samuel R. and {Howlett}, Cullan and {Calcino}, Josh},
        title = "{Can redshift errors bias measurements of the Hubble Constant?}",
      journal = {\mnras},
         year = 2019,
        month = dec,
       volume = {490},
       number = {2},
        pages = {2948-2957},
          doi = {10.1093/mnras/stz2652},
archivePrefix = {arXiv},
       eprint = {1907.12639},
 primaryClass = {astro-ph.CO},
       adsurl = {https://ui.adsabs.harvard.edu/abs/2019MNRAS.490.2948D}
}

@ARTICLE{2013arXiv1308.0847L,
       author = {{Levi}, Michael and {Bebek}, Chris and {Beers}, Timothy and {Blum}, Robert and {Cahn}, Robert and {Eisenstein}, Daniel and {Flaugher}, Brenna and {Honscheid}, Klaus and {Kron}, Richard and {Lahav}, Ofer and {McDonald}, Patrick and {Roe}, Natalie and {Schlegel}, David and {representing the DESI collaboration}},
        title = "{The DESI Experiment, a whitepaper for Snowmass 2013}",
      journal = {arXiv e-prints},
         year = 2013,
        month = aug,
          eid = {arXiv:1308.0847},
        pages = {arXiv:1308.0847},
          doi = {10.48550/arXiv.1308.0847},
archivePrefix = {arXiv},
       eprint = {1308.0847},
 primaryClass = {astro-ph.CO},
       adsurl = {https://ui.adsabs.harvard.edu/abs/2013arXiv1308.0847L}
}

@ARTICLE{2016arXiv161100036D,
       author = {{DESI Collaboration} and {Aghamousa}, Amir and {Aguilar}, Jessica and {Ahlen}, Steve and {Alam}, Shadab and {Allen}, Lori E. and {Allende Prieto}, Carlos and {Annis}, James and {Bailey}, Stephen and {Balland}, Christophe and {Ballester}, Otger and {Baltay}, Charles and {Beaufore}, Lucas and {Bebek}, Chris and {Beers}, Timothy C. and {Bell}, Eric F. and {Bernal}, Jos{\'e} Luis and {Besuner}, Robert and {Beutler}, Florian and {Blake}, Chris and {Bleuler}, Hannes and {Blomqvist}, Michael and {Blum}, Robert and {Bolton}, Adam S. and {Briceno}, Cesar and {Brooks}, David and {Brownstein}, Joel R. and {Buckley-Geer}, Elizabeth and {Burden}, Angela and {Burtin}, Etienne and {Busca}, Nicolas G. and {Cahn}, Robert N. and {Cai}, Yan-Chuan and {Cardiel-Sas}, Laia and {Carlberg}, Raymond G. and {Carton}, Pierre-Henri and {Casas}, Ricard and {Castander}, Francisco J. and {Cervantes-Cota}, Jorge L. and {Claybaugh}, Todd M. and {Close}, Madeline and {Coker}, Carl T. and {Cole}, Shaun and {Comparat}, Johan and {Cooper}, Andrew P. and {Cousinou}, M. -C. and {Crocce}, Martin and {Cuby}, Jean-Gabriel and {Cunningham}, Daniel P. and {Davis}, Tamara M. and {Dawson}, Kyle S. and {de la Macorra}, Axel and {De Vicente}, Juan and {Delubac}, Timoth{\'e}e and {Derwent}, Mark and {Dey}, Arjun and {Dhungana}, Govinda and {Ding}, Zhejie and {Doel}, Peter and {Duan}, Yutong T. and {Ealet}, Anne and {Edelstein}, Jerry and {Eftekharzadeh}, Sarah and {Eisenstein}, Daniel J. and {Elliott}, Ann and {Escoffier}, St{\'e}phanie and {Evatt}, Matthew and {Fagrelius}, Parker and {Fan}, Xiaohui and {Fanning}, Kevin and {Farahi}, Arya and {Farihi}, Jay and {Favole}, Ginevra and {Feng}, Yu and {Fernandez}, Enrique and {Findlay}, Joseph R. and {Finkbeiner}, Douglas P. and {Fitzpatrick}, Michael J. and {Flaugher}, Brenna and {Flender}, Samuel and {Font-Ribera}, Andreu and {Forero-Romero}, Jaime E. and {Fosalba}, Pablo and {Frenk}, Carlos S. and {Fumagalli}, Michele and {Gaensicke}, Boris T. and {Gallo}, Giuseppe and {Garcia-Bellido}, Juan and {Gaztanaga}, Enrique and {Pietro Gentile Fusillo}, Nicola and {Gerard}, Terry and {Gershkovich}, Irena and {Giannantonio}, Tommaso and {Gillet}, Denis and {Gonzalez-de-Rivera}, Guillermo and {Gonzalez-Perez}, Violeta and {Gott}, Shelby and {Graur}, Or and {Gutierrez}, Gaston and {Guy}, Julien and {Habib}, Salman and {Heetderks}, Henry and {Heetderks}, Ian and {Heitmann}, Katrin and {Hellwing}, Wojciech A. and {Herrera}, David A. and {Ho}, Shirley and {Holland}, Stephen and {Honscheid}, Klaus and {Huff}, Eric and {Hutchinson}, Timothy A. and {Huterer}, Dragan and {Hwang}, Ho Seong and {Illa Laguna}, Joseph Maria and {Ishikawa}, Yuzo and {Jacobs}, Dianna and {Jeffrey}, Niall and {Jelinsky}, Patrick and {Jennings}, Elise and {Jiang}, Linhua and {Jimenez}, Jorge and {Johnson}, Jennifer and {Joyce}, Richard and {Jullo}, Eric and {Juneau}, St{\'e}phanie and {Kama}, Sami and {Karcher}, Armin and {Karkar}, Sonia and {Kehoe}, Robert and {Kennamer}, Noble and {Kent}, Stephen and {Kilbinger}, Martin and {Kim}, Alex G. and {Kirkby}, David and {Kisner}, Theodore and {Kitanidis}, Ellie and {Kneib}, Jean-Paul and {Koposov}, Sergey and {Kovacs}, Eve and {Koyama}, Kazuya and {Kremin}, Anthony and {Kron}, Richard and {Kronig}, Luzius and {Kueter-Young}, Andrea and {Lacey}, Cedric G. and {Lafever}, Robin and {Lahav}, Ofer and {Lambert}, Andrew and {Lampton}, Michael and {Landriau}, Martin and {Lang}, Dustin and {Lauer}, Tod R. and {Le Goff}, Jean-Marc and {Le Guillou}, Laurent and {Le Van Suu}, Auguste and {Lee}, Jae Hyeon and {Lee}, Su-Jeong and {Leitner}, Daniela and {Lesser}, Michael and {Levi}, Michael E. and {L'Huillier}, Benjamin and {Li}, Baojiu and {Liang}, Ming and {Lin}, Huan and {Linder}, Eric and {Loebman}, Sarah R. and {Luki{\'c}}, Zarija and {Ma}, Jun and {MacCrann}, Niall and {Magneville}, Christophe and {Makarem}, Laleh and {Manera}, Marc and {Manser}, Christopher J. and {Marshall}, Robert and {Martini}, Paul and {Massey}, Richard and {Matheson}, Thomas and {McCauley}, Jeremy and {McDonald}, Patrick and {McGreer}, Ian D. and {Meisner}, Aaron and {Metcalfe}, Nigel and {Miller}, Timothy N. and {Miquel}, Ramon and {Moustakas}, John and {Myers}, Adam and {Naik}, Milind and {Newman}, Jeffrey A. and {Nichol}, Robert C. and {Nicola}, Andrina and {Nicolati da Costa}, Luiz and {Nie}, Jundan and {Niz}, Gustavo and {Norberg}, Peder and {Nord}, Brian and {Norman}, Dara and {Nugent}, Peter and {O'Brien}, Thomas and {Oh}, Minji and {Olsen}, Knut A.~G. and {Padilla}, Cristobal and {Padmanabhan}, Hamsa and {Padmanabhan}, Nikhil and {Palanque-Delabrouille}, Nathalie and {Palmese}, Antonella and {Pappalardo}, Daniel and {P{\^a}ris}, Isabelle and {Park}, Changbom and {Patej}, Anna and {Peacock}, John A. and {Peiris}, Hiranya V. and {Peng}, Xiyan and {Percival}, Will J. and {Perruchot}, Sandrine and {Pieri}, Matthew M. and {Pogge}, Richard and {Pollack}, Jennifer E. and {Poppett}, Claire and {Prada}, Francisco and {Prakash}, Abhishek and {Probst}, Ronald G. and {Rabinowitz}, David and {Raichoor}, Anand and {Ree}, Chang Hee and {Refregier}, Alexandre and {Regal}, Xavier and {Reid}, Beth and {Reil}, Kevin and {Rezaie}, Mehdi and {Rockosi}, Constance M. and {Roe}, Natalie and {Ronayette}, Samuel and {Roodman}, Aaron and {Ross}, Ashley J. and {Ross}, Nicholas P. and {Rossi}, Graziano and {Rozo}, Eduardo and {Ruhlmann-Kleider}, Vanina and {Rykoff}, Eli S. and {Sabiu}, Cristiano and {Samushia}, Lado and {Sanchez}, Eusebio and {Sanchez}, Javier and {Schlegel}, David J. and {Schneider}, Michael and {Schubnell}, Michael and {Secroun}, Aur{\'e}lia and {Seljak}, Uros and {Seo}, Hee-Jong and {Serrano}, Santiago and {Shafieloo}, Arman and {Shan}, Huanyuan and {Sharples}, Ray and {Sholl}, Michael J. and {Shourt}, William V. and {Silber}, Joseph H. and {Silva}, David R. and {Sirk}, Martin M. and {Slosar}, Anze and {Smith}, Alex and {Smoot}, George F. and {Som}, Debopam and {Song}, Yong-Seon and {Sprayberry}, David and {Staten}, Ryan and {Stefanik}, Andy and {Tarle}, Gregory and {Sien Tie}, Suk and {Tinker}, Jeremy L. and {Tojeiro}, Rita and {Valdes}, Francisco and {Valenzuela}, Octavio and {Valluri}, Monica and {Vargas-Magana}, Mariana and {Verde}, Licia and {Walker}, Alistair R. and {Wang}, Jiali and {Wang}, Yuting and {Weaver}, Benjamin A. and {Weaverdyck}, Curtis and {Wechsler}, Risa H. and {Weinberg}, David H. and {White}, Martin and {Yang}, Qian and {Yeche}, Christophe and {Zhang}, Tianmeng and {Zhao}, Gong-Bo and {Zheng}, Yi and {Zhou}, Xu and {Zhou}, Zhimin and {Zhu}, Yaling and {Zou}, Hu and {Zu}, Ying},
        title = "{The DESI Experiment Part I: Science,Targeting, and Survey Design}",
      journal = {arXiv e-prints},
         year = 2016,
        month = oct,
          eid = {arXiv:1611.00036},
        pages = {arXiv:1611.00036},
          doi = {10.48550/arXiv.1611.00036},
archivePrefix = {arXiv},
       eprint = {1611.00036},
 primaryClass = {astro-ph.IM},
       adsurl = {https://ui.adsabs.harvard.edu/abs/2016arXiv161100036D}
}

@ARTICLE{2016arXiv161100037D,
       author = {{DESI Collaboration} and {Aghamousa}, Amir and {Aguilar}, Jessica and {Ahlen}, Steve and {Alam}, Shadab and {Allen}, Lori E. and {Allende Prieto}, Carlos and {Annis}, James and {Bailey}, Stephen and {Balland}, Christophe and {Ballester}, Otger and {Baltay}, Charles and {Beaufore}, Lucas and {Bebek}, Chris and {Beers}, Timothy C. and {Bell}, Eric F. and {Bernal}, Jos{\'e} Luis and {Besuner}, Robert and {Beutler}, Florian and {Blake}, Chris and {Bleuler}, Hannes and {Blomqvist}, Michael and {Blum}, Robert and {Bolton}, Adam S. and {Briceno}, Cesar and {Brooks}, David and {Brownstein}, Joel R. and {Buckley-Geer}, Elizabeth and {Burden}, Angela and {Burtin}, Etienne and {Busca}, Nicolas G. and {Cahn}, Robert N. and {Cai}, Yan-Chuan and {Cardiel-Sas}, Laia and {Carlberg}, Raymond G. and {Carton}, Pierre-Henri and {Casas}, Ricard and {Castander}, Francisco J. and {Cervantes-Cota}, Jorge L. and {Claybaugh}, Todd M. and {Close}, Madeline and {Coker}, Carl T. and {Cole}, Shaun and {Comparat}, Johan and {Cooper}, Andrew P. and {Cousinou}, M. -C. and {Crocce}, Martin and {Cuby}, Jean-Gabriel and {Cunningham}, Daniel P. and {Davis}, Tamara M. and {Dawson}, Kyle S. and {de la Macorra}, Axel and {De Vicente}, Juan and {Delubac}, Timoth{\'e}e and {Derwent}, Mark and {Dey}, Arjun and {Dhungana}, Govinda and {Ding}, Zhejie and {Doel}, Peter and {Duan}, Yutong T. and {Ealet}, Anne and {Edelstein}, Jerry and {Eftekharzadeh}, Sarah and {Eisenstein}, Daniel J. and {Elliott}, Ann and {Escoffier}, St{\'e}phanie and {Evatt}, Matthew and {Fagrelius}, Parker and {Fan}, Xiaohui and {Fanning}, Kevin and {Farahi}, Arya and {Farihi}, Jay and {Favole}, Ginevra and {Feng}, Yu and {Fernandez}, Enrique and {Findlay}, Joseph R. and {Finkbeiner}, Douglas P. and {Fitzpatrick}, Michael J. and {Flaugher}, Brenna and {Flender}, Samuel and {Font-Ribera}, Andreu and {Forero-Romero}, Jaime E. and {Fosalba}, Pablo and {Frenk}, Carlos S. and {Fumagalli}, Michele and {Gaensicke}, Boris T. and {Gallo}, Giuseppe and {Garcia-Bellido}, Juan and {Gaztanaga}, Enrique and {Pietro Gentile Fusillo}, Nicola and {Gerard}, Terry and {Gershkovich}, Irena and {Giannantonio}, Tommaso and {Gillet}, Denis and {Gonzalez-de-Rivera}, Guillermo and {Gonzalez-Perez}, Violeta and {Gott}, Shelby and {Graur}, Or and {Gutierrez}, Gaston and {Guy}, Julien and {Habib}, Salman and {Heetderks}, Henry and {Heetderks}, Ian and {Heitmann}, Katrin and {Hellwing}, Wojciech A. and {Herrera}, David A. and {Ho}, Shirley and {Holland}, Stephen and {Honscheid}, Klaus and {Huff}, Eric and {Hutchinson}, Timothy A. and {Huterer}, Dragan and {Hwang}, Ho Seong and {Illa Laguna}, Joseph Maria and {Ishikawa}, Yuzo and {Jacobs}, Dianna and {Jeffrey}, Niall and {Jelinsky}, Patrick and {Jennings}, Elise and {Jiang}, Linhua and {Jimenez}, Jorge and {Johnson}, Jennifer and {Joyce}, Richard and {Jullo}, Eric and {Juneau}, St{\'e}phanie and {Kama}, Sami and {Karcher}, Armin and {Karkar}, Sonia and {Kehoe}, Robert and {Kennamer}, Noble and {Kent}, Stephen and {Kilbinger}, Martin and {Kim}, Alex G. and {Kirkby}, David and {Kisner}, Theodore and {Kitanidis}, Ellie and {Kneib}, Jean-Paul and {Koposov}, Sergey and {Kovacs}, Eve and {Koyama}, Kazuya and {Kremin}, Anthony and {Kron}, Richard and {Kronig}, Luzius and {Kueter-Young}, Andrea and {Lacey}, Cedric G. and {Lafever}, Robin and {Lahav}, Ofer and {Lambert}, Andrew and {Lampton}, Michael and {Landriau}, Martin and {Lang}, Dustin and {Lauer}, Tod R. and {Le Goff}, Jean-Marc and {Le Guillou}, Laurent and {Le Van Suu}, Auguste and {Lee}, Jae Hyeon and {Lee}, Su-Jeong and {Leitner}, Daniela and {Lesser}, Michael and {Levi}, Michael E. and {L'Huillier}, Benjamin and {Li}, Baojiu and {Liang}, Ming and {Lin}, Huan and {Linder}, Eric and {Loebman}, Sarah R. and {Luki{\'c}}, Zarija and {Ma}, Jun and {MacCrann}, Niall and {Magneville}, Christophe and {Makarem}, Laleh and {Manera}, Marc and {Manser}, Christopher J. and {Marshall}, Robert and {Martini}, Paul and {Massey}, Richard and {Matheson}, Thomas and {McCauley}, Jeremy and {McDonald}, Patrick and {McGreer}, Ian D. and {Meisner}, Aaron and {Metcalfe}, Nigel and {Miller}, Timothy N. and {Miquel}, Ramon and {Moustakas}, John and {Myers}, Adam and {Naik}, Milind and {Newman}, Jeffrey A. and {Nichol}, Robert C. and {Nicola}, Andrina and {Nicolati da Costa}, Luiz and {Nie}, Jundan and {Niz}, Gustavo and {Norberg}, Peder and {Nord}, Brian and {Norman}, Dara and {Nugent}, Peter and {O'Brien}, Thomas and {Oh}, Minji and {Olsen}, Knut A.~G. and {Padilla}, Cristobal and {Padmanabhan}, Hamsa and {Padmanabhan}, Nikhil and {Palanque-Delabrouille}, Nathalie and {Palmese}, Antonella and {Pappalardo}, Daniel and {P{\^a}ris}, Isabelle and {Park}, Changbom and {Patej}, Anna and {Peacock}, John A. and {Peiris}, Hiranya V. and {Peng}, Xiyan and {Percival}, Will J. and {Perruchot}, Sandrine and {Pieri}, Matthew M. and {Pogge}, Richard and {Pollack}, Jennifer E. and {Poppett}, Claire and {Prada}, Francisco and {Prakash}, Abhishek and {Probst}, Ronald G. and {Rabinowitz}, David and {Raichoor}, Anand and {Ree}, Chang Hee and {Refregier}, Alexandre and {Regal}, Xavier and {Reid}, Beth and {Reil}, Kevin and {Rezaie}, Mehdi and {Rockosi}, Constance M. and {Roe}, Natalie and {Ronayette}, Samuel and {Roodman}, Aaron and {Ross}, Ashley J. and {Ross}, Nicholas P. and {Rossi}, Graziano and {Rozo}, Eduardo and {Ruhlmann-Kleider}, Vanina and {Rykoff}, Eli S. and {Sabiu}, Cristiano and {Samushia}, Lado and {Sanchez}, Eusebio and {Sanchez}, Javier and {Schlegel}, David J. and {Schneider}, Michael and {Schubnell}, Michael and {Secroun}, Aur{\'e}lia and {Seljak}, Uros and {Seo}, Hee-Jong and {Serrano}, Santiago and {Shafieloo}, Arman and {Shan}, Huanyuan and {Sharples}, Ray and {Sholl}, Michael J. and {Shourt}, William V. and {Silber}, Joseph H. and {Silva}, David R. and {Sirk}, Martin M. and {Slosar}, Anze and {Smith}, Alex and {Smoot}, George F. and {Som}, Debopam and {Song}, Yong-Seon and {Sprayberry}, David and {Staten}, Ryan and {Stefanik}, Andy and {Tarle}, Gregory and {Sien Tie}, Suk and {Tinker}, Jeremy L. and {Tojeiro}, Rita and {Valdes}, Francisco and {Valenzuela}, Octavio and {Valluri}, Monica and {Vargas-Magana}, Mariana and {Verde}, Licia and {Walker}, Alistair R. and {Wang}, Jiali and {Wang}, Yuting and {Weaver}, Benjamin A. and {Weaverdyck}, Curtis and {Wechsler}, Risa H. and {Weinberg}, David H. and {White}, Martin and {Yang}, Qian and {Yeche}, Christophe and {Zhang}, Tianmeng and {Zhao}, Gong-Bo and {Zheng}, Yi and {Zhou}, Xu and {Zhou}, Zhimin and {Zhu}, Yaling and {Zou}, Hu and {Zu}, Ying},
        title = "{The DESI Experiment Part II: Instrument Design}",
      journal = {arXiv e-prints},
         year = 2016,
        month = oct,
          eid = {arXiv:1611.00037},
        pages = {arXiv:1611.00037},
          doi = {10.48550/arXiv.1611.00037},
archivePrefix = {arXiv},
       eprint = {1611.00037},
 primaryClass = {astro-ph.IM},
       adsurl = {https://ui.adsabs.harvard.edu/abs/2016arXiv161100037D}
}

@ARTICLE{2019AJ....157..168D,
       author = {{Dey}, Arjun and {Schlegel}, David J. and {Lang}, Dustin and {Blum}, Robert and {Burleigh}, Kaylan and {Fan}, Xiaohui and {Findlay}, Joseph R. and {Finkbeiner}, Doug and {Herrera}, David and {Juneau}, St{\'e}phanie and {Landriau}, Martin and {Levi}, Michael and {McGreer}, Ian and {Meisner}, Aaron and {Myers}, Adam D. and {Moustakas}, John and {Nugent}, Peter and {Patej}, Anna and {Schlafly}, Edward F. and {Walker}, Alistair R. and {Valdes}, Francisco and {Weaver}, Benjamin A. and {Y{\`e}che}, Christophe and {Zou}, Hu and {Zhou}, Xu and {Abareshi}, Behzad and {Abbott}, T.~M.~C. and {Abolfathi}, Bela and {Aguilera}, C. and {Alam}, Shadab and {Allen}, Lori and {Alvarez}, A. and {Annis}, James and {Ansarinejad}, Behzad and {Aubert}, Marie and {Beechert}, Jacqueline and {Bell}, Eric F. and {BenZvi}, Segev Y. and {Beutler}, Florian and {Bielby}, Richard M. and {Bolton}, Adam S. and {Brice{\~n}o}, C{\'e}sar and {Buckley-Geer}, Elizabeth J. and {Butler}, Karen and {Calamida}, Annalisa and {Carlberg}, Raymond G. and {Carter}, Paul and {Casas}, Ricard and {Castander}, Francisco J. and {Choi}, Yumi and {Comparat}, Johan and {Cukanovaite}, Elena and {Delubac}, Timoth{\'e}e and {DeVries}, Kaitlin and {Dey}, Sharmila and {Dhungana}, Govinda and {Dickinson}, Mark and {Ding}, Zhejie and {Donaldson}, John B. and {Duan}, Yutong and {Duckworth}, Christopher J. and {Eftekharzadeh}, Sarah and {Eisenstein}, Daniel J. and {Etourneau}, Thomas and {Fagrelius}, Parker A. and {Farihi}, Jay and {Fitzpatrick}, Mike and {Font-Ribera}, Andreu and {Fulmer}, Leah and {G{\"a}nsicke}, Boris T. and {Gaztanaga}, Enrique and {George}, Koshy and {Gerdes}, David W. and {Gontcho}, Satya Gontcho A. and {Gorgoni}, Claudio and {Green}, Gregory and {Guy}, Julien and {Harmer}, Diane and {Hernandez}, M. and {Honscheid}, Klaus and {Huang}, Lijuan Wendy and {James}, David J. and {Jannuzi}, Buell T. and {Jiang}, Linhua and {Joyce}, Richard and {Karcher}, Armin and {Karkar}, Sonia and {Kehoe}, Robert and {Kneib}, Jean-Paul and {Kueter-Young}, Andrea and {Lan}, Ting-Wen and {Lauer}, Tod R. and {Le Guillou}, Laurent and {Le Van Suu}, Auguste and {Lee}, Jae Hyeon and {Lesser}, Michael and {Perreault Levasseur}, Laurence and {Li}, Ting S. and {Mann}, Justin L. and {Marshall}, Robert and {Mart{\'\i}nez-V{\'a}zquez}, C.~E. and {Martini}, Paul and {du Mas des Bourboux}, H{\'e}lion and {McManus}, Sean and {Meier}, Tobias Gabriel and {M{\'e}nard}, Brice and {Metcalfe}, Nigel and {Mu{\~n}oz-Guti{\'e}rrez}, Andrea and {Najita}, Joan and {Napier}, Kevin and {Narayan}, Gautham and {Newman}, Jeffrey A. and {Nie}, Jundan and {Nord}, Brian and {Norman}, Dara J. and {Olsen}, Knut A.~G. and {Paat}, Anthony and {Palanque-Delabrouille}, Nathalie and {Peng}, Xiyan and {Poppett}, Claire L. and {Poremba}, Megan R. and {Prakash}, Abhishek and {Rabinowitz}, David and {Raichoor}, Anand and {Rezaie}, Mehdi and {Robertson}, A.~N. and {Roe}, Natalie A. and {Ross}, Ashley J. and {Ross}, Nicholas P. and {Rudnick}, Gregory and {Safonova}, Sasha and {Saha}, Abhijit and {S{\'a}nchez}, F. Javier and {Savary}, Elodie and {Schweiker}, Heidi and {Scott}, Adam and {Seo}, Hee-Jong and {Shan}, Huanyuan and {Silva}, David R. and {Slepian}, Zachary and {Soto}, Christian and {Sprayberry}, David and {Staten}, Ryan and {Stillman}, Coley M. and {Stupak}, Robert J. and {Summers}, David L. and {Sien Tie}, Suk and {Tirado}, H. and {Vargas-Maga{\~n}a}, Mariana and {Vivas}, A. Katherina and {Wechsler}, Risa H. and {Williams}, Doug and {Yang}, Jinyi and {Yang}, Qian and {Yapici}, Tolga and {Zaritsky}, Dennis and {Zenteno}, A. and {Zhang}, Kai and {Zhang}, Tianmeng and {Zhou}, Rongpu and {Zhou}, Zhimin},
        title = "{Overview of the DESI Legacy Imaging Surveys}",
      journal = {\aj},
         year = 2019,
        month = may,
       volume = {157},
       number = {5},
          eid = {168},
        pages = {168},
          doi = {10.3847/1538-3881/ab089d},
archivePrefix = {arXiv},
       eprint = {1804.08657},
 primaryClass = {astro-ph.IM},
       adsurl = {https://ui.adsabs.harvard.edu/abs/2019AJ....157..168D}
}

@ARTICLE{2023AJ....165....9S,
       author = {{Silber}, Joseph Harry and {Fagrelius}, Parker and {Fanning}, Kevin and {Schubnell}, Michael and {Aguilar}, Jessica Nicole and {Ahlen}, Steven and {Ameel}, Jon and {Ballester}, Otger and {Baltay}, Charles and {Bebek}, Chris and {Benton Beard}, Dominic and {Besuner}, Robert and {Cardiel-Sas}, Laia and {Casas}, Ricard and {Castander}, Francisco Javier and {Claybaugh}, Todd and {Dobson}, Carl and {Duan}, Yutong and {Dunlop}, Patrick and {Edelstein}, Jerry and {Emmet}, William T. and {Elliott}, Ann and {Evatt}, Matthew and {Gershkovich}, Irena and {Guy}, Julien and {Harris}, Stu and {Heetderks}, Henry and {Heetderks}, Ian and {Honscheid}, Klaus and {Illa}, Jose Maria and {Jelinsky}, Patrick and {Jelinsky}, Sharon R. and {Jimenez}, Jorge and {Karcher}, Armin and {Kent}, Stephen and {Kirkby}, David and {Kneib}, Jean-Paul and {Lambert}, Andrew and {Lampton}, Mike and {Leitner}, Daniela and {Levi}, Michael and {McCauley}, Jeremy and {Meisner}, Aaron and {Miller}, Timothy N. and {Miquel}, Ramon and {Mundet}, Juli{\'a} and {Poppett}, Claire and {Rabinowitz}, David and {Reil}, Kevin and {Roman}, David and {Schlegel}, David and {Serrano}, Santiago and {Van Shourt}, William and {Sprayberry}, David and {Tarl{\'e}}, Gregory and {Tie}, Suk Sien and {Weaverdyck}, Curtis and {Zhang}, Kai and {Azzaro}, Marco and {Bailey}, Stephen and {Becerril}, Santiago and {Blackwell}, Tami and {Bouri}, Mohamed and {Brooks}, David and {Buckley-Geer}, Elizabeth and {Castro}, Jose Pe{\~n}ate and {Derwent}, Mark and {Dey}, Arjun and {Dhungana}, Govinda and {Doel}, Peter and {Eisenstein}, Daniel J. and {Fahim}, Nasib and {Garcia-Bellido}, Juan and {Gazta{\~n}aga}, Enrique and {A Gontcho}, Satya Gontcho and {Gutierrez}, Gaston and {H{\"o}rler}, Philipp and {Kehoe}, Robert and {Kisner}, Theodore and {Kremin}, Anthony and {Kronig}, Luzius and {Landriau}, Martin and {Le Guillou}, Laurent and {Martini}, Paul and {Moustakas}, John and {Palanque-Delabrouille}, Nathalie and {Peng}, Xiyan and {Percival}, Will and {Prada}, Francisco and {Allende Prieto}, Carlos and {de Rivera}, Guillermo Gonzalez and {Sanchez}, Eusebio and {Sanchez}, Justo and {Sharples}, Ray and {Soares-Santos}, Marcelle and {Schlafly}, Edward and {Weaver}, Benjamin Alan and {Zhou}, Zhimin and {Zhu}, Yaling and {Zou}, Hu and {DESI Collaboration}},
        title = "{The Robotic Multiobject Focal Plane System of the Dark Energy Spectroscopic Instrument (DESI)}",
      journal = {\aj},
         year = 2023,
        month = jan,
       volume = {165},
       number = {1},
          eid = {9},
        pages = {9},
          doi = {10.3847/1538-3881/ac9ab1},
archivePrefix = {arXiv},
       eprint = {2205.09014},
 primaryClass = {astro-ph.IM},
       adsurl = {https://ui.adsabs.harvard.edu/abs/2023AJ....165....9S}
}

@ARTICLE{2023AJ....165..144G,
       author = {{Guy}, J. and {Bailey}, S. and {Kremin}, A. and {Alam}, Shadab and {Alexander}, D.~M. and {Allende Prieto}, C. and {BenZvi}, S. and {Bolton}, A.~S. and {Brooks}, D. and {Chaussidon}, E. and {Cooper}, A.~P. and {Dawson}, K. and {de la Macorra}, A. and {Dey}, A. and {Dey}, Biprateep and {Dhungana}, G. and {Eisenstein}, D.~J. and {Font-Ribera}, A. and {Forero-Romero}, J.~E. and {Gazta{\~n}aga}, E. and {Gontcho A Gontcho}, S. and {Green}, D. and {Honscheid}, K. and {Ishak}, M. and {Kehoe}, R. and {Kirkby}, D. and {Kisner}, T. and {Koposov}, Sergey E. and {Lan}, Ting-Wen and {Landriau}, M. and {Le Guillou}, L. and {Levi}, Michael E. and {Magneville}, C. and {Manser}, Christopher J. and {Martini}, P. and {Meisner}, Aaron M. and {Miquel}, R. and {Moustakas}, J. and {Myers}, Adam D. and {Newman}, Jeffrey A. and {Nie}, Jundan and {Palanque-Delabrouille}, N. and {Percival}, W.~J. and {Poppett}, C. and {Prada}, F. and {Raichoor}, A. and {Ravoux}, C. and {Ross}, A.~J. and {Schlafly}, E.~F. and {Schlegel}, D. and {Schubnell}, M. and {Sharples}, Ray M. and {Tarl{\'e}}, Gregory and {Weaver}, B.~A. and {Y{\'e}che}, Christophe and {Zhou}, Rongpu and {Zhou}, Zhimin and {Zou}, H.},
        title = "{The Spectroscopic Data Processing Pipeline for the Dark Energy Spectroscopic Instrument}",
      journal = {\aj},
         year = 2023,
        month = apr,
       volume = {165},
       number = {4},
          eid = {144},
        pages = {144},
          doi = {10.3847/1538-3881/acb212},
archivePrefix = {arXiv},
       eprint = {2209.14482},
 primaryClass = {astro-ph.IM},
       adsurl = {https://ui.adsabs.harvard.edu/abs/2023AJ....165..144G}
}

@ARTICLE{2023AJ....166..259S,
       author = {{Schlafly}, Edward F. and {Kirkby}, David and {Schlegel}, David J. and {Myers}, Adam D. and {Raichoor}, Anand and {Dawson}, Kyle and {Aguilar}, Jessica and {Allende Prieto}, Carlos and {Bailey}, Stephen and {BenZvi}, Segev and {Bermejo-Climent}, Jose and {Brooks}, David and {de la Macorra}, Axel and {Dey}, Arjun and {Doel}, Peter and {Fanning}, Kevin and {Font-Ribera}, Andreu and {Forero-Romero}, Jaime E. and {Garc{\'\i}a-Bellido}, Juan and {Gontcho A Gontcho}, Satya and {Guy}, Julien and {Hahn}, ChangHoon and {Honscheid}, Klaus and {Ishak}, Mustapha and {Juneau}, St{\'e}phanie and {Kehoe}, Robert and {Kisner}, Theodore and {Kremin}, Anthony and {Landriau}, Martin and {Lang}, Dustin A. and {Lasker}, James and {Levi}, Michael E. and {Magneville}, Christophe and {Manser}, Christopher J. and {Martini}, Paul and {Meisner}, Aaron M. and {Miquel}, Ramon and {Moustakas}, John and {Newman}, Jeffrey A. and {Nie}, Jundan and {Palanque-Delabrouille}, Nathalie. and {Percival}, Will J. and {Poppett}, Claire and {Rockosi}, Constance and {Ross}, Ashley J. and {Rossi}, Graziano and {Tarl{\'e}}, Gregory and {Weaver}, Benjamin A. and {Y{\`e}che}, Christophe and {Zhou}, Rongpu and {DESI Collaboration}},
        title = "{Survey Operations for the Dark Energy Spectroscopic Instrument}",
      journal = {\aj},
         year = 2023,
        month = dec,
       volume = {166},
       number = {6},
          eid = {259},
        pages = {259},
          doi = {10.3847/1538-3881/ad0832},
archivePrefix = {arXiv},
       eprint = {2306.06309},
 primaryClass = {astro-ph.CO},
       adsurl = {https://ui.adsabs.harvard.edu/abs/2023AJ....166..259S}
}

@ARTICLE{2024AJ....167...62D,
       author = {{DESI Collaboration} and {Adame}, A.~G. and {Aguilar}, J. and {Ahlen}, S. and {Alam}, S. and {Aldering}, G. and {Alexander}, D.~M. and {Alfarsy}, R. and {Allende Prieto}, C. and {Alvarez}, M. and {Alves}, O. and {Anand}, A. and {Andrade-Oliveira}, F. and {Armengaud}, E. and {Asorey}, J. and {Avila}, S. and {Aviles}, A. and {Bailey}, S. and {Balaguera-Antol{\'\i}nez}, A. and {Ballester}, O. and {Baltay}, C. and {Bault}, A. and {Bautista}, J. and {Behera}, J. and {Beltran}, S.~F. and {BenZvi}, S. and {Beraldo e Silva}, L. and {Bermejo-Climent}, J.~R. and {Berti}, A. and {Besuner}, R. and {Beutler}, F. and {Bianchi}, D. and {Blake}, C. and {Blum}, R. and {Bolton}, A.~S. and {Brieden}, S. and {Brodzeller}, A. and {Brooks}, D. and {Brown}, Z. and {Buckley-Geer}, E. and {Burtin}, E. and {Cabayol-Garcia}, L. and {Cai}, Z. and {Canning}, R. and {Cardiel-Sas}, L. and {Carnero Rosell}, A. and {Castander}, F.~J. and {Cervantes-Cota}, J.~L. and {Chabanier}, S. and {Chaussidon}, E. and {Chaves-Montero}, J. and {Chen}, S. and {Chen}, X. and {Chuang}, C. and {Claybaugh}, T. and {Cole}, S. and {Cooper}, A.~P. and {Cuceu}, A. and {Davis}, T.~M. and {Dawson}, K. and {de Belsunce}, R. and {de la Cruz}, R. and {de la Macorra}, A. and {de Mattia}, A. and {Demina}, R. and {Demirbozan}, U. and {DeRose}, J. and {Dey}, A. and {Dey}, B. and {Dhungana}, G. and {Ding}, J. and {Ding}, Z. and {Doel}, P. and {Doshi}, R. and {Douglass}, K. and {Edge}, A. and {Eftekharzadeh}, S. and {Eisenstein}, D.~J. and {Elliott}, A. and {Escoffier}, S. and {Fagrelius}, P. and {Fan}, X. and {Fanning}, K. and {Fawcett}, V.~A. and {Ferraro}, S. and {Ereza}, J. and {Flaugher}, B. and {Font-Ribera}, A. and {Forero-S{\'a}nchez}, D. and {Forero-Romero}, J.~E. and {Frenk}, C.~S. and {G{\"a}nsicke}, B.~T. and {Garc{\'\i}a}, L. {\'A}. and {Garc{\'\i}a-Bellido}, J. and {Garcia-Quintero}, C. and {Garrison}, L.~H. and {Gil-Mar{\'\i}n}, H. and {Golden-Marx}, J. and {Gontcho A Gontcho}, S. and {Gonzalez-Morales}, A.~X. and {Gonzalez-Perez}, V. and {Gordon}, C. and {Graur}, O. and {Green}, D. and {Gruen}, D. and {Guy}, J. and {Hadzhiyska}, B. and {Hahn}, C. and {Han}, J.~J. and {Hanif}, M.~M.~S. and {Herrera-Alcantar}, H.~K. and {Honscheid}, K. and {Hou}, J. and {Howlett}, C. and {Huterer}, D. and {Ir{\v{s}}i{\v{c}}}, V. and {Ishak}, M. and {Jana}, A. and {Jiang}, L. and {Jimenez}, J. and {Jing}, Y.~P. and {Joudaki}, S. and {Jullo}, E. and {Joyce}, R. and {Juneau}, S. and {Kizhuprakkat}, N. and {Kara{\c{c}}ayl{\i}}, N.~G. and {Karim}, T. and {Kehoe}, R. and {Kent}, S. and {Khederlarian}, A. and {Kim}, S. and {Kirkby}, D. and {Kisner}, T. and {Kitaura}, F. and {Kneib}, J. and {Koposov}, S.~E. and {Kov{\'a}cs}, A. and {Kremin}, A. and {Krolewski}, A. and {L'Huillier}, B. and {Lahav}, O. and {Lambert}, A. and {Lamman}, C. and {Lan}, T. -W. and {Landriau}, M. and {Lang}, D. and {Lange}, J.~U. and {Lasker}, J. and {Le Guillou}, L. and {Leauthaud}, A. and {Levi}, M.~E. and {Li}, T.~S. and {Linder}, E. and {Lyons}, A. and {Magneville}, C. and {Manera}, M. and {Manser}, C.~J. and {Margala}, D. and {Martini}, P. and {McDonald}, P. and {Medina}, G.~E. and {Medina-Varela}, L. and {Meisner}, A. and {Mena-Fern{\'a}ndez}, J. and {Meneses-Rizo}, J. and {Mezcua}, M. and {Miquel}, R. and {Montero-Camacho}, P. and {Moon}, J. and {Moore}, S. and {Moustakas}, J. and {Mueller}, E. and {Mundet}, J. and {Mu{\~n}oz-Guti{\'e}rrez}, A. and {Myers}, A.~D. and {Nadathur}, S. and {Napolitano}, L. and {Neveux}, R. and {Newman}, J.~A. and {Nie}, J. and {Niz}, G. and {Norberg}, P. and {Noriega}, H.~E. and {Paillas}, E. and {Palanque-Delabrouille}, N. and {Palmese}, A. and {Zhiwei}, P. and {Parkinson}, D. and {Penmetsa}, S. and {Percival}, W.~J. and {P{\'e}rez-Fern{\'a}ndez}, A. and {P{\'e}rez-R{\`a}fols}, I. and {Pieri}, M. and {Poppett}, C. and {Porredon}, A. and {Prada}, F. and {Pucha}, R. and {Raichoor}, A. and {Ram{\'\i}rez-P{\'e}rez}, C. and {Ramirez-Solano}, S. and {Rashkovetskyi}, M. and {Ravoux}, C. and {Rocher}, A. and {Rockosi}, C. and {Ross}, A.~J. and {Rossi}, G. and {Ruggeri}, R. and {Ruhlmann-Kleider}, V. and {Sabiu}, C.~G. and {Said}, K. and {Saintonge}, A. and {Samushia}, L. and {Sanchez}, E. and {Saulder}, C. and {Schaan}, E. and {Schlafly}, E.~F. and {Schlegel}, D. and {Scholte}, D. and {Schubnell}, M. and {Seo}, H. and {Shafieloo}, A. and {Sharples}, R. and {Sheu}, W. and {Silber}, J. and {Sinigaglia}, F. and {Siudek}, M. and {Slepian}, Z. and {Smith}, A. and {Sprayberry}, D. and {Stephey}, L. and {Su{\'a}rez-P{\'e}rez}, J. and {Sun}, Z. and {Tan}, T. and {Tarl{\'e}}, G. and {Tojeiro}, R. and {Ure{\~n}a-L{\'o}pez}, L.~A. and {Vaisakh}, R. and {Valcin}, D. and {Valdes}, F. and {Valluri}, M. and {Vargas-Maga{\~n}a}, M. and {Variu}, A. and {Verde}, L. and {Walther}, M. and {Wang}, B. and {Wang}, M.~S. and {Weaver}, B.~A. and {Weaverdyck}, N. and {Wechsler}, R.~H. and {White}, M. and {Xie}, Y. and {Yang}, J. and {Y{\`e}che}, C. and {Yu}, J. and {Yuan}, S. and {Zhang}, H. and {Zhang}, Z. and {Zhao}, C. and {Zheng}, Z. and {Zhou}, R. and {Zhou}, Z. and {Zou}, H. and {Zou}, S. and {Zu}, Y. and {DESI Collaboration}},
        title = "{Validation of the Scientific Program for the Dark Energy Spectroscopic Instrument}",
      journal = {\aj},
         year = 2024,
        month = feb,
       volume = {167},
       number = {2},
          eid = {62},
        pages = {62},
          doi = {10.3847/1538-3881/ad0b08},
archivePrefix = {arXiv},
       eprint = {2306.06307},
 primaryClass = {astro-ph.CO},
       adsurl = {https://ui.adsabs.harvard.edu/abs/2024AJ....167...62D}
}

@ARTICLE{2024AJ....168...58D,
       author = {{DESI Collaboration} and {Adame}, A.~G. and {Aguilar}, J. and {Ahlen}, S. and {Alam}, S. and {Aldering}, G. and {Alexander}, D.~M. and {Alfarsy}, R. and {Allende Prieto}, C. and {Alvarez}, M. and {Alves}, O. and {Anand}, A. and {Andrade-Oliveira}, F. and {Armengaud}, E. and {Asorey}, J. and {Avila}, S. and {Aviles}, A. and {Bailey}, S. and {Balaguera-Antol{\'\i}nez}, A. and {Ballester}, O. and {Baltay}, C. and {Bault}, A. and {Bautista}, J. and {Behera}, J. and {Beltran}, S.~F. and {BenZvi}, S. and {Beraldo e Silva}, L. and {Bermejo-Climent}, J.~R. and {Berti}, A. and {Besuner}, R. and {Beutler}, F. and {Bianchi}, D. and {Blake}, C. and {Blum}, R. and {Bolton}, A.~S. and {Brieden}, S. and {Brodzeller}, A. and {Brooks}, D. and {Brown}, Z. and {Buckley-Geer}, E. and {Burtin}, E. and {Cabayol-Garcia}, L. and {Cai}, Z. and {Canning}, R. and {Cardiel-Sas}, L. and {Carnero Rosell}, A. and {Castander}, F.~J. and {Cervantes-Cota}, J.~L. and {Chabanier}, S. and {Chaussidon}, E. and {Chaves-Montero}, J. and {Chen}, S. and {Chen}, X. and {Chuang}, C. and {Claybaugh}, T. and {Cole}, S. and {Cooper}, A.~P. and {Cuceu}, A. and {Davis}, T.~M. and {Dawson}, K. and {de Belsunce}, R. and {de la Cruz}, R. and {de la Macorra}, A. and {Della Costa}, J. and {de Mattia}, A. and {Demina}, R. and {Demirbozan}, U. and {DeRose}, J. and {Dey}, A. and {Dey}, B. and {Dhungana}, G. and {Ding}, J. and {Ding}, Z. and {Doel}, P. and {Doshi}, R. and {Douglass}, K. and {Edge}, A. and {Eftekharzadeh}, S. and {Eisenstein}, D.~J. and {Elliott}, A. and {Ereza}, J. and {Escoffier}, S. and {Fagrelius}, P. and {Fan}, X. and {Fanning}, K. and {Fawcett}, V.~A. and {Ferraro}, S. and {Flaugher}, B. and {Font-Ribera}, A. and {Forero-Romero}, J.~E. and {Forero-S{\'a}nchez}, D. and {Frenk}, C.~S. and {G{\"a}nsicke}, B.~T. and {Garc{\'\i}a}, L. {\'A}. and {Garc{\'\i}a-Bellido}, J. and {Garcia-Quintero}, C. and {Garrison}, L.~H. and {Gil-Mar{\'\i}n}, H. and {Golden-Marx}, J. and {Gontcho A Gontcho}, S. and {Gonzalez-Morales}, A.~X. and {Gonzalez-Perez}, V. and {Gordon}, C. and {Graur}, O. and {Green}, D. and {Gruen}, D. and {Guy}, J. and {Hadzhiyska}, B. and {Hahn}, C. and {Han}, J.~J. and {Hanif}, M.~M.~S. and {Herrera-Alcantar}, H.~K. and {Honscheid}, K. and {Hou}, J. and {Howlett}, C. and {Huterer}, D. and {Ir{\v{s}}i{\v{c}}}, V. and {Ishak}, M. and {Jacques}, A. and {Jana}, A. and {Jiang}, L. and {Jimenez}, J. and {Jing}, Y.~P. and {Joudaki}, S. and {Joyce}, R. and {Jullo}, E. and {Juneau}, S. and {Kara{\c{c}}ayl{\i}}, N.~G. and {Karim}, T. and {Kehoe}, R. and {Kent}, S. and {Khederlarian}, A. and {Kim}, S. and {Kirkby}, D. and {Kisner}, T. and {Kitaura}, F. and {Kizhuprakkat}, N. and {Kneib}, J. and {Koposov}, S.~E. and {Kov{\'a}cs}, A. and {Kremin}, A. and {Krolewski}, A. and {L'Huillier}, B. and {Lahav}, O. and {Lambert}, A. and {Lamman}, C. and {Lan}, T. -W. and {Landriau}, M. and {Lang}, D. and {Lange}, J.~U. and {Lasker}, J. and {Leauthaud}, A. and {Le Guillou}, L. and {Levi}, M.~E. and {Li}, T.~S. and {Linder}, E. and {Lyons}, A. and {Magneville}, C. and {Manera}, M. and {Manser}, C.~J. and {Margala}, D. and {Martini}, P. and {McDonald}, P. and {Medina}, G.~E. and {Medina-Varela}, L. and {Meisner}, A. and {Mena-Fern{\'a}ndez}, J. and {Meneses-Rizo}, J. and {Mezcua}, M. and {Miquel}, R. and {Montero-Camacho}, P. and {Moon}, J. and {Moore}, S. and {Moustakas}, J. and {Mueller}, E. and {Mundet}, J. and {Mu{\~n}oz-Guti{\'e}rrez}, A. and {Myers}, A.~D. and {Nadathur}, S. and {Napolitano}, L. and {Neveux}, R. and {Newman}, J.~A. and {Nie}, J. and {Nikutta}, R. and {Niz}, G. and {Norberg}, P. and {Noriega}, H.~E. and {Paillas}, E. and {Palanque-Delabrouille}, N. and {Palmese}, A. and {Pan}, Z. and {Parkinson}, D. and {Penmetsa}, S. and {Percival}, W.~J. and {P{\'e}rez-Fern{\'a}ndez}, A. and {P{\'e}rez-R{\`a}fols}, I. and {Pieri}, M. and {Poppett}, C. and {Porredon}, A. and {Pothier}, S. and {Prada}, F. and {Pucha}, R. and {Raichoor}, A. and {Ram{\'\i}rez-P{\'e}rez}, C. and {Ramirez-Solano}, S. and {Rashkovetskyi}, M. and {Ravoux}, C. and {Rocher}, A. and {Rockosi}, C. and {Ross}, A.~J. and {Rossi}, G. and {Ruggeri}, R. and {Ruhlmann-Kleider}, V. and {Sabiu}, C.~G. and {Said}, K. and {Saintonge}, A. and {Samushia}, L. and {Sanchez}, E. and {Saulder}, C. and {Schaan}, E. and {Schlafly}, E.~F. and {Schlegel}, D. and {Scholte}, D. and {Schubnell}, M. and {Seo}, H. and {Shafieloo}, A. and {Sharples}, R. and {Sheu}, W. and {Silber}, J. and {Sinigaglia}, F. and {Siudek}, M. and {Slepian}, Z. and {Smith}, A. and {Soumagnac}, M.~T. and {Sprayberry}, D. and {Stephey}, L. and {Su{\'a}rez-P{\'e}rez}, J. and {Sun}, Z. and {Tan}, T. and {Tarl{\'e}}, G. and {Tojeiro}, R. and {Ure{\~n}a-L{\'o}pez}, L.~A. and {Vaisakh}, R. and {Valcin}, D. and {Valdes}, F. and {Valluri}, M. and {Vargas-Maga{\~n}a}, M. and {Variu}, A. and {Verde}, L. and {Walther}, M. and {Wang}, B. and {Wang}, M.~S. and {Weaver}, B.~A. and {Weaverdyck}, N. and {Wechsler}, R.~H. and {White}, M. and {Xie}, Y. and {Yang}, J. and {Y{\`e}che}, C. and {Yu}, J. and {Yuan}, S. and {Zhang}, H. and {Zhang}, Z. and {Zhao}, C. and {Zheng}, Z. and {Zhou}, R. and {Zhou}, Z. and {Zou}, H. and {Zou}, S. and {Zu}, Y.},
        title = "{The Early Data Release of the Dark Energy Spectroscopic Instrument}",
      journal = {\aj},
         year = 2024,
        month = aug,
       volume = {168},
       number = {2},
          eid = {58},
        pages = {58},
          doi = {10.3847/1538-3881/ad3217},
archivePrefix = {arXiv},
       eprint = {2306.06308},
 primaryClass = {astro-ph.CO},
       adsurl = {https://ui.adsabs.harvard.edu/abs/2024AJ....168...58D}
}

@ARTICLE{2023AJ....165..253H,
       author = {{Hahn}, ChangHoon and {Wilson}, Michael J. and {Ruiz-Macias}, Omar and {Cole}, Shaun and {Weinberg}, David H. and {Moustakas}, John and {Kremin}, Anthony and {Tinker}, Jeremy L. and {Smith}, Alex and {Wechsler}, Risa H. and {Ahlen}, Steven and {Alam}, Shadab and {Bailey}, Stephen and {Brooks}, David and {Cooper}, Andrew P. and {Davis}, Tamara M. and {Dawson}, Kyle and {Dey}, Arjun and {Dey}, Biprateep and {Eftekharzadeh}, Sarah and {Eisenstein}, Daniel J. and {Fanning}, Kevin and {Forero-Romero}, Jaime E. and {Frenk}, Carlos S. and {Gazta{\~n}aga}, Enrique and {A Gontcho}, Satya Gontcho and {Guy}, Julien and {Honscheid}, Klaus and {Ishak}, Mustapha and {Juneau}, St{\'e}phanie and {Kehoe}, Robert and {Kisner}, Theodore and {Lan}, Ting-Wen and {Landriau}, Martin and {Le Guillou}, Laurent and {Levi}, Michael E. and {Magneville}, Christophe and {Martini}, Paul and {Meisner}, Aaron and {Myers}, Adam D. and {Nie}, Jundan and {Norberg}, Peder and {Palanque-Delabrouille}, Nathalie and {Percival}, Will J. and {Poppett}, Claire and {Prada}, Francisco and {Raichoor}, Anand and {Ross}, Ashley J. and {Gaines}, Sasha and {Saulder}, Christoph and {Schlafly}, Eddie and {Schlegel}, David and {Sierra-Porta}, David and {Tarle}, Gregory and {Weaver}, Benjamin A. and {Y{\`e}che}, Christophe and {Zarrouk}, Pauline and {Zhou}, Rongpu and {Zhou}, Zhimin and {Zou}, Hu},
        title = "{The DESI Bright Galaxy Survey: Final Target Selection, Design, and Validation}",
      journal = {\aj},
         year = 2023,
        month = jun,
       volume = {165},
       number = {6},
          eid = {253},
        pages = {253},
          doi = {10.3847/1538-3881/accff8},
archivePrefix = {arXiv},
       eprint = {2208.08512},
 primaryClass = {astro-ph.CO},
       adsurl = {https://ui.adsabs.harvard.edu/abs/2023AJ....165..253H}
}

@ARTICLE{2022ApJ...938..113S,
       author = {{Scolnic}, Dan and {Brout}, Dillon and {Carr}, Anthony and {Riess}, Adam G. and {Davis}, Tamara M. and {Dwomoh}, Arianna and {Jones}, David O. and {Ali}, Noor and {Charvu}, Pranav and {Chen}, Rebecca and {Peterson}, Erik R. and {Popovic}, Brodie and {Rose}, Benjamin M. and {Wood}, Charlotte M. and {Brown}, Peter J. and {Chambers}, Ken and {Coulter}, David A. and {Dettman}, Kyle G. and {Dimitriadis}, Georgios and {Filippenko}, Alexei V. and {Foley}, Ryan J. and {Jha}, Saurabh W. and {Kilpatrick}, Charles D. and {Kirshner}, Robert P. and {Pan}, Yen-Chen and {Rest}, Armin and {Rojas-Bravo}, Cesar and {Siebert}, Matthew R. and {Stahl}, Benjamin E. and {Zheng}, WeiKang},
        title = "{The Pantheon+ Analysis: The Full Data Set and Light-curve Release}",
      journal = {\apj},
         year = 2022,
        month = oct,
       volume = {938},
       number = {2},
          eid = {113},
        pages = {113},
          doi = {10.3847/1538-4357/ac8b7a},
archivePrefix = {arXiv},
       eprint = {2112.03863},
 primaryClass = {astro-ph.CO},
       adsurl = {https://ui.adsabs.harvard.edu/abs/2022ApJ...938..113S}
}

@ARTICLE{2022ApJ...938..111B,
       author = {{Brout}, Dillon and {Taylor}, Georgie and {Scolnic}, Dan and {Wood}, Charlotte M. and {Rose}, Benjamin M. and {Vincenzi}, Maria and {Dwomoh}, Arianna and {Lidman}, Christopher and {Riess}, Adam and {Ali}, Noor and {Qu}, Helen and {Dai}, Mi},
        title = "{The Pantheon+ Analysis: SuperCal-fragilistic Cross Calibration, Retrained SALT2 Light-curve Model, and Calibration Systematic Uncertainty}",
      journal = {\apj},
         year = 2022,
        month = oct,
       volume = {938},
       number = {2},
          eid = {111},
        pages = {111},
          doi = {10.3847/1538-4357/ac8bcc},
archivePrefix = {arXiv},
       eprint = {2112.03864},
 primaryClass = {astro-ph.CO},
       adsurl = {https://ui.adsabs.harvard.edu/abs/2022ApJ...938..111B}
}

@ARTICLE{2025JCAP...04..012A,
       author = {{DESI Collaboration} and {Adame}, A.~G. and {Aguilar}, J. and {Ahlen}, S. and {Alam}, S. and {Alexander}, D.~M. and {Alvarez}, M. and {Alves}, O. and {Anand}, A. and {Andrade}, U. and {Armengaud}, E. and {Avila}, S. and {Aviles}, A. and {Awan}, H. and {Bailey}, S. and {Baltay}, C. and {Bault}, A. and {Behera}, J. and {BenZvi}, S. and {Beutler}, F. and {Bianchi}, D. and {Blake}, C. and {Blum}, R. and {Brieden}, S. and {Brodzeller}, A. and {Brooks}, D. and {Buckley-Geer}, E. and {Burtin}, E. and {Calderon}, R. and {Canning}, R. and {Carnero Rosell}, A. and {Cereskaite}, R. and {Cervantes-Cota}, J.~L. and {Chabanier}, S. and {Chaussidon}, E. and {Chaves-Montero}, J. and {Chen}, S. and {Chen}, X. and {Claybaugh}, T. and {Cole}, S. and {Cuceu}, A. and {Davis}, T.~M. and {Dawson}, K. and {de la Macorra}, A. and {de Mattia}, A. and {Deiosso}, N. and {Dey}, A. and {Dey}, B. and {Ding}, Z. and {Doel}, P. and {Edelstein}, J. and {Eftekharzadeh}, S. and {Eisenstein}, D.~J. and {Elliott}, A. and {Fagrelius}, P. and {Fanning}, K. and {Ferraro}, S. and {Ereza}, J. and {Findlay}, N. and {Flaugher}, B. and {Font-Ribera}, A. and {Forero-S{\'a}nchez}, D. and {Forero-Romero}, J.~E. and {Garcia-Quintero}, C. and {Gazta{\~n}aga}, E. and {Gil-Mar{\'\i}n}, H. and {Gontcho a Gontcho}, S. and {Gonzalez-Morales}, A.~X. and {Gonzalez-Perez}, V. and {Gordon}, C. and {Green}, D. and {Gruen}, D. and {Gsponer}, R. and {Gutierrez}, G. and {Guy}, J. and {Hadzhiyska}, B. and {Hahn}, C. and {Hanif}, M.~M.~S. and {Herrera-Alcantar}, H.~K. and {Honscheid}, K. and {Howlett}, C. and {Huterer}, D. and {Ir{\v{s}}i{\v{c}}}, V. and {Ishak}, M. and {Juneau}, S. and {Kara{\c{c}}ayl{\i}}, N.~G. and {Kehoe}, R. and {Kent}, S. and {Kirkby}, D. and {Kong}, H. and {Kremin}, A. and {Krolewski}, A. and {Lai}, Y. and {Lan}, T. -W. and {Landriau}, M. and {Lang}, D. and {Lasker}, J. and {Le Goff}, J.~M. and {Le Guillou}, L. and {Leauthaud}, A. and {Levi}, M.~E. and {Li}, T.~S. and {Linder}, E. and {Lodha}, K. and {Magneville}, C. and {Manera}, M. and {Margala}, D. and {Martini}, P. and {Maus}, M. and {McDonald}, P. and {Medina-Varela}, L. and {Meisner}, A. and {Mena-Fern{\'a}ndez}, J. and {Miquel}, R. and {Moon}, J. and {Moore}, S. and {Moustakas}, J. and {Mueller}, E. and {Mu{\~n}oz-Guti{\'e}rrez}, A. and {Myers}, A.~D. and {Nadathur}, S. and {Napolitano}, L. and {Neveux}, R. and {Newman}, J.~A. and {Nguyen}, N.~M. and {Nie}, J. and {Niz}, G. and {Noriega}, H.~E. and {Padmanabhan}, N. and {Paillas}, E. and {Palanque-Delabrouille}, N. and {Pan}, J. and {Penmetsa}, S. and {Percival}, W.~J. and {Pieri}, M.~M. and {Pinon}, M. and {Poppett}, C. and {Porredon}, A. and {Prada}, F. and {P{\'e}rez-Fern{\'a}ndez}, A. and {P{\'e}rez-R{\`a}fols}, I. and {Rabinowitz}, D. and {Raichoor}, A. and {Ram{\'\i}rez-P{\'e}rez}, C. and {Ramirez-Solano}, S. and {Rashkovetskyi}, M. and {Ravoux}, C. and {Rezaie}, M. and {Rich}, J. and {Rocher}, A. and {Rockosi}, C. and {Roe}, N.~A. and {Rosado-Marin}, A. and {Ross}, A.~J. and {Rossi}, G. and {Ruggeri}, R. and {Ruhlmann-Kleider}, V. and {Samushia}, L. and {Sanchez}, E. and {Saulder}, C. and {Schlafly}, E.~F. and {Schlegel}, D. and {Schubnell}, M. and {Seo}, H. and {Sharples}, R. and {Silber}, J. and {Slosar}, A. and {Smith}, A. and {Sprayberry}, D. and {Swanson}, J. and {Tan}, T. and {Tarl{\'e}}, G. and {Trusov}, S. and {Vaisakh}, R. and {Valcin}, D. and {Valdes}, F. and {Vargas-Maga{\~n}a}, M. and {Verde}, L. and {Walther}, M. and {Wang}, B. and {Wang}, M.~S. and {Weaver}, B.~A. and {Weaverdyck}, N. and {Wechsler}, R.~H. and {Weinberg}, D.~H. and {White}, M. and {Wilson}, M.~J. and {Yu}, J. and {Yu}, Y. and {Yuan}, S. and {Y{\`e}che}, C. and {Zaborowski}, E.~A. and {Zarrouk}, P. and {Zhang}, H. and {Zhao}, C. and {Zhao}, R. and {Zhou}, R. and {Zou}, H. and {DESI Collaboration}},
        title = "{DESI 2024 III: baryon acoustic oscillations from galaxies and quasars}",
      journal = {\jcap},
         year = 2025,
        month = apr,
       volume = {2025},
       number = {4},
          eid = {012},
        pages = {012},
          doi = {10.1088/1475-7516/2025/04/012},
archivePrefix = {arXiv},
       eprint = {2404.03000},
 primaryClass = {astro-ph.CO},
       adsurl = {https://ui.adsabs.harvard.edu/abs/2025JCAP...04..012A}
}

@ARTICLE{2010A&A...523A...7G,
       author = {{Guy}, J. and {Sullivan}, M. and {Conley}, A. and {Regnault}, N. and {Astier}, P. and {Balland}, C. and {Basa}, S. and {Carlberg}, R.~G. and {Fouchez}, D. and {Hardin}, D. and {Hook}, I.~M. and {Howell}, D.~A. and {Pain}, R. and {Palanque-Delabrouille}, N. and {Perrett}, K.~M. and {Pritchet}, C.~J. and {Rich}, J. and {Ruhlmann-Kleider}, V. and {Balam}, D. and {Baumont}, S. and {Ellis}, R.~S. and {Fabbro}, S. and {Fakhouri}, H.~K. and {Fourmanoit}, N. and {Gonz{\'a}lez-Gait{\'a}n}, S. and {Graham}, M.~L. and {Hsiao}, E. and {Kronborg}, T. and {Lidman}, C. and {Mourao}, A.~M. and {Perlmutter}, S. and {Ripoche}, P. and {Suzuki}, N. and {Walker}, E.~S.},
        title = "{The Supernova Legacy Survey 3-year sample: Type Ia supernovae photometric distances and cosmological constraints}",
      journal = {\aap},
         year = 2010,
        month = nov,
       volume = {523},
          eid = {A7},
        pages = {A7},
          doi = {10.1051/0004-6361/201014468},
archivePrefix = {arXiv},
       eprint = {1010.4743},
 primaryClass = {astro-ph.CO},
       adsurl = {https://ui.adsabs.harvard.edu/abs/2010A&A...523A...7G}
}

@ARTICLE{1998A&A...331..815T,
       author = {{Tripp}, Robert},
        title = "{A two-parameter luminosity correction for Type IA supernovae}",
      journal = {\aap},
         year = 1998,
        month = mar,
       volume = {331},
        pages = {815-820},
       adsurl = {https://ui.adsabs.harvard.edu/abs/1998A&A...331..815T}
}

@ARTICLE{2017ApJ...836...56K,
       author = {{Kessler}, R. and {Scolnic}, D.},
        title = "{Correcting Type Ia Supernova Distances for Selection Biases and Contamination in Photometrically Identified Samples}",
      journal = {\apj},
         year = 2017,
        month = feb,
       volume = {836},
       number = {1},
          eid = {56},
        pages = {56},
          doi = {10.3847/1538-4357/836/1/56},
archivePrefix = {arXiv},
       eprint = {1610.04677},
 primaryClass = {astro-ph.CO},
       adsurl = {https://ui.adsabs.harvard.edu/abs/2017ApJ...836...56K}
}

@ARTICLE{2021ApJ...913...49P,
       author = {{Popovic}, Brodie and {Brout}, Dillon and {Kessler}, Richard and {Scolnic}, Dan and {Lu}, Lisa},
        title = "{Improved Treatment of Host-galaxy Correlations in Cosmological Analyses with Type Ia Supernovae}",
      journal = {\apj},
         year = 2021,
        month = may,
       volume = {913},
       number = {1},
          eid = {49},
        pages = {49},
          doi = {10.3847/1538-4357/abf14f},
archivePrefix = {arXiv},
       eprint = {2102.01776},
 primaryClass = {astro-ph.CO},
       adsurl = {https://ui.adsabs.harvard.edu/abs/2021ApJ...913...49P}
}

@ARTICLE{2022ApJ...938..110B,
       author = {{Brout}, Dillon and {Scolnic}, Dan and {Popovic}, Brodie and {Riess}, Adam G. and {Carr}, Anthony and {Zuntz}, Joe and {Kessler}, Rick and {Davis}, Tamara M. and {Hinton}, Samuel and {Jones}, David and {Kenworthy}, W. D'Arcy and {Peterson}, Erik R. and {Said}, Khaled and {Taylor}, Georgie and {Ali}, Noor and {Armstrong}, Patrick and {Charvu}, Pranav and {Dwomoh}, Arianna and {Meldorf}, Cole and {Palmese}, Antonella and {Qu}, Helen and {Rose}, Benjamin M. and {Sanchez}, Bruno and {Stubbs}, Christopher W. and {Vincenzi}, Maria and {Wood}, Charlotte M. and {Brown}, Peter J. and {Chen}, Rebecca and {Chambers}, Ken and {Coulter}, David A. and {Dai}, Mi and {Dimitriadis}, Georgios and {Filippenko}, Alexei V. and {Foley}, Ryan J. and {Jha}, Saurabh W. and {Kelsey}, Lisa and {Kirshner}, Robert P. and {M{\"o}ller}, Anais and {Muir}, Jessie and {Nadathur}, Seshadri and {Pan}, Yen-Chen and {Rest}, Armin and {Rojas-Bravo}, Cesar and {Sako}, Masao and {Siebert}, Matthew R. and {Smith}, Mat and {Stahl}, Benjamin E. and {Wiseman}, Phil},
        title = "{The Pantheon+ Analysis: Cosmological Constraints}",
      journal = {\apj},
         year = 2022,
        month = oct,
       volume = {938},
       number = {2},
          eid = {110},
        pages = {110},
          doi = {10.3847/1538-4357/ac8e04},
archivePrefix = {arXiv},
       eprint = {2202.04077},
 primaryClass = {astro-ph.CO},
       adsurl = {https://ui.adsabs.harvard.edu/abs/2022ApJ...938..110B}
}

@ARTICLE{betoule2014,
       author = {{Betoule}, M. and {Kessler}, R. and {Guy}, J. and {Mosher}, J. and {Hardin}, D. and {Biswas}, R. and {Astier}, P. and {El-Hage}, P. and {Konig}, M. and {Kuhlmann}, S. and {Marriner}, J. and {Pain}, R. and {Regnault}, N. and {Balland}, C. and {Bassett}, B.~A. and {Brown}, P.~J. and {Campbell}, H. and {Carlberg}, R.~G. and {Cellier-Holzem}, F. and {Cinabro}, D. and {Conley}, A. and {D'Andrea}, C.~B. and {DePoy}, D.~L. and {Doi}, M. and {Ellis}, R.~S. and {Fabbro}, S. and {Filippenko}, A.~V. and {Foley}, R.~J. and {Frieman}, J.~A. and {Fouchez}, D. and {Galbany}, L. and {Goobar}, A. and {Gupta}, R.~R. and {Hill}, G.~J. and {Hlozek}, R. and {Hogan}, C.~J. and {Hook}, I.~M. and {Howell}, D.~A. and {Jha}, S.~W. and {Le Guillou}, L. and {Leloudas}, G. and {Lidman}, C. and {Marshall}, J.~L. and {M{\"o}ller}, A. and {Mour{\~a}o}, A.~M. and {Neveu}, J. and {Nichol}, R. and {Olmstead}, M.~D. and {Palanque-Delabrouille}, N. and {Perlmutter}, S. and {Prieto}, J.~L. and {Pritchet}, C.~J. and {Richmond}, M. and {Riess}, A.~G. and {Ruhlmann-Kleider}, V. and {Sako}, M. and {Schahmaneche}, K. and {Schneider}, D.~P. and {Smith}, M. and {Sollerman}, J. and {Sullivan}, M. and {Walton}, N.~A. and {Wheeler}, C.~J.},
        title = "{Improved cosmological constraints from a joint analysis of the SDSS-II and SNLS supernova samples}",
      journal = {\aap},
         year = 2014,
        month = aug,
       volume = {568},
          eid = {A22},
        pages = {A22},
          doi = {10.1051/0004-6361/201423413},
archivePrefix = {arXiv},
       eprint = {1401.4064},
 primaryClass = {astro-ph.CO},
       adsurl = {https://ui.adsabs.harvard.edu/abs/2014A&A...568A..22B}
}

@ARTICLE{nicolas2021,
       author = {{Nicolas}, N. and {Rigault}, M. and {Copin}, Y. and {Graziani}, R. and {Aldering}, G. and {Briday}, M. and {Kim}, Y. -L. and {Nordin}, J. and {Perlmutter}, S. and {Smith}, M.},
        title = "{Redshift evolution of the underlying type Ia supernova stretch distribution}",
      journal = {\aap},
         year = 2021,
        month = may,
       volume = {649},
          eid = {A74},
        pages = {A74},
          doi = {10.1051/0004-6361/202038447},
archivePrefix = {arXiv},
       eprint = {2005.09441},
 primaryClass = {astro-ph.CO},
       adsurl = {https://ui.adsabs.harvard.edu/abs/2021A&A...649A..74N}
}

@ARTICLE{rigault2020,
       author = {{Rigault}, M. and {Brinnel}, V. and {Aldering}, G. and {Antilogus}, P. and {Aragon}, C. and {Bailey}, S. and {Baltay}, C. and {Barbary}, K. and {Bongard}, S. and {Boone}, K. and {Buton}, C. and {Childress}, M. and {Chotard}, N. and {Copin}, Y. and {Dixon}, S. and {Fagrelius}, P. and {Feindt}, U. and {Fouchez}, D. and {Gangler}, E. and {Hayden}, B. and {Hillebrandt}, W. and {Howell}, D.~A. and {Kim}, A. and {Kowalski}, M. and {Kuesters}, D. and {Leget}, P. -F. and {Lombardo}, S. and {Lin}, Q. and {Nordin}, J. and {Pain}, R. and {Pecontal}, E. and {Pereira}, R. and {Perlmutter}, S. and {Rabinowitz}, D. and {Runge}, K. and {Rubin}, D. and {Saunders}, C. and {Smadja}, G. and {Sofiatti}, C. and {Suzuki}, N. and {Taubenberger}, S. and {Tao}, C. and {Thomas}, R.~C.},
        title = "{Strong dependence of Type Ia supernova standardization on the local specific star formation rate}",
      journal = {\aap},
         year = 2020,
        month = dec,
       volume = {644},
          eid = {A176},
        pages = {A176},
          doi = {10.1051/0004-6361/201730404},
archivePrefix = {arXiv},
       eprint = {1806.03849},
 primaryClass = {astro-ph.CO},
       adsurl = {https://ui.adsabs.harvard.edu/abs/2020A&A...644A.176R}
}

@ARTICLE{smithGeneratingMockGalaxy2024,
       author = {{Smith}, A. and {Grove}, C. and {Cole}, S. and {Norberg}, P. and {Zarrouk}, P. and {Yuan}, S. and {Aguilar}, J. and {Ahlen}, S. and {Brooks}, D. and {Claybaugh}, T. and {de la Macorra}, A. and {Doel}, P. and {Forero-Romero}, J.~E. and {Gazta{\~n}aga}, E. and {Gontcho}, S. Gontcho A. and {Hahn}, C. and {Kehoe}, R. and {Kremin}, A. and {Levi}, M.~E. and {Manera}, M. and {Meisner}, A. and {Miquel}, R. and {Moustakas}, J. and {Nie}, J. and {Percival}, W.~J. and {Rezaie}, M. and {Rossi}, G. and {Sanchez}, E. and {Seo}, H. and {Tarl{\'e}}, G. and {Zhou}, Z.},
        title = "{Generating mock galaxy catalogues for flux-limited samples like the DESI Bright Galaxy Survey}",
      journal = {\mnras},
         year = 2024,
        month = jul,
       volume = {532},
       number = {1},
        pages = {903-919},
          doi = {10.1093/mnras/stae1503},
archivePrefix = {arXiv},
       eprint = {2312.08792},
 primaryClass = {astro-ph.CO},
       adsurl = {https://ui.adsabs.harvard.edu/abs/2024MNRAS.532..903S}
}

@ARTICLE{garrisonABACUSCosmologicalNbody2021,
       author = {{Garrison}, Lehman H. and {Eisenstein}, Daniel J. and {Ferrer}, Douglas and {Maksimova}, Nina A. and {Pinto}, Philip A.},
        title = "{The ABACUS cosmological N-body code}",
      journal = {\mnras},
         year = 2021,
        month = nov,
       volume = {508},
       number = {1},
        pages = {575-596},
          doi = {10.1093/mnras/stab2482},
archivePrefix = {arXiv},
       eprint = {2110.11392},
 primaryClass = {astro-ph.CO},
       adsurl = {https://ui.adsabs.harvard.edu/abs/2021MNRAS.508..575G}
}

@ARTICLE{maksimovaABACUSSUMMITMassiveSet2021,
       author = {{Maksimova}, Nina A. and {Garrison}, Lehman H. and {Eisenstein}, Daniel J. and {Hadzhiyska}, Boryana and {Bose}, Sownak and {Satterthwaite}, Thomas P.},
        title = "{ABACUSSUMMIT: a massive set of high-accuracy, high-resolution N-body simulations}",
      journal = {\mnras},
         year = 2021,
        month = dec,
       volume = {508},
       number = {3},
        pages = {4017-4037},
          doi = {10.1093/mnras/stab2484},
archivePrefix = {arXiv},
       eprint = {2110.11398},
 primaryClass = {astro-ph.CO},
       adsurl = {https://ui.adsabs.harvard.edu/abs/2021MNRAS.508.4017M}
}

@ARTICLE{hadzhiyskaCOMPASONewHalo2022,
       author = {{Hadzhiyska}, Boryana and {Eisenstein}, Daniel and {Bose}, Sownak and {Garrison}, Lehman H. and {Maksimova}, Nina},
        title = "{COMPASO: A new halo finder for competitive assignment to spherical overdensities}",
      journal = {\mnras},
         year = 2022,
        month = jan,
       volume = {509},
       number = {1},
        pages = {501-521},
          doi = {10.1093/mnras/stab2980},
archivePrefix = {arXiv},
       eprint = {2110.11408},
 primaryClass = {astro-ph.CO},
       adsurl = {https://ui.adsabs.harvard.edu/abs/2022MNRAS.509..501H}
}

@ARTICLE{smithLightconeCatalogueMillenniumXXL2017,
       author = {{Smith}, Alex and {Cole}, Shaun and {Baugh}, Carlton and {Zheng}, Zheng and {Angulo}, Ra{\'u}l and {Norberg}, Peder and {Zehavi}, Idit},
        title = "{A lightcone catalogue from the Millennium-XXL simulation}",
      journal = {\mnras},
         year = 2017,
        month = oct,
       volume = {470},
       number = {4},
        pages = {4646-4661},
          doi = {10.1093/mnras/stx1432},
archivePrefix = {arXiv},
       eprint = {1701.06581},
 primaryClass = {astro-ph.CO},
       adsurl = {https://ui.adsabs.harvard.edu/abs/2017MNRAS.470.4646S}
}

@ARTICLE{smithLightconeCatalogueMillenniumXXL2022,
       author = {{Smith}, Alex and {Cole}, Shaun and {Grove}, Cameron and {Norberg}, Peder and {Zarrouk}, Pauline},
        title = "{A light-cone catalogue from the Millennium-XXL simulation: improved spatial interpolation and colour distributions for the DESI BGS}",
      journal = {\mnras},
         year = 2022,
        month = nov,
       volume = {516},
       number = {3},
        pages = {4529-4542},
          doi = {10.1093/mnras/stac2519},
archivePrefix = {arXiv},
       eprint = {2207.04902},
 primaryClass = {astro-ph.CO},
       adsurl = {https://ui.adsabs.harvard.edu/abs/2022MNRAS.516.4529S}
}

@ARTICLE{sullivan_rates_2006,
       author = {{Sullivan}, M. and {Le Borgne}, D. and {Pritchet}, C.~J. and {Hodsman}, A. and {Neill}, J.~D. and {Howell}, D.~A. and {Carlberg}, R.~G. and {Astier}, P. and {Aubourg}, E. and {Balam}, D. and {Basa}, S. and {Conley}, A. and {Fabbro}, S. and {Fouchez}, D. and {Guy}, J. and {Hook}, I. and {Pain}, R. and {Palanque-Delabrouille}, N. and {Perrett}, K. and {Regnault}, N. and {Rich}, J. and {Taillet}, R. and {Baumont}, S. and {Bronder}, J. and {Ellis}, R.~S. and {Filiol}, M. and {Lusset}, V. and {Perlmutter}, S. and {Ripoche}, P. and {Tao}, C.},
        title = "{Rates and Properties of Type Ia Supernovae as a Function of Mass and Star Formation in Their Host Galaxies}",
      journal = {\apj},
         year = 2006,
        month = sep,
       volume = {648},
       number = {2},
        pages = {868-883},
          doi = {10.1086/506137},
archivePrefix = {arXiv},
       eprint = {astro-ph/0605455},
 primaryClass = {astro-ph},
       adsurl = {https://ui.adsabs.harvard.edu/abs/2006ApJ...648..868S}
}

@ARTICLE{mannucci_supernova_2005,
       author = {{Mannucci}, F. and {Della Valle}, M. and {Panagia}, N. and {Cappellaro}, E. and {Cresci}, G. and {Maiolino}, R. and {Petrosian}, A. and {Turatto}, M.},
        title = "{The supernova rate per unit mass}",
      journal = {\aap},
         year = 2005,
        month = apr,
       volume = {433},
       number = {3},
        pages = {807-814},
          doi = {10.1051/0004-6361:20041411},
archivePrefix = {arXiv},
       eprint = {astro-ph/0411450},
 primaryClass = {astro-ph},
       adsurl = {https://ui.adsabs.harvard.edu/abs/2005A&A...433..807M}
}

@ARTICLE{2024MNRAS.527..501B,
       author = {{Blake}, Chris and {Turner}, Ryan J.},
        title = "{On the correlations of galaxy peculiar velocities and their covariance}",
      journal = {\mnras},
         year = 2024,
        month = jan,
       volume = {527},
       number = {1},
        pages = {501-520},
          doi = {10.1093/mnras/stad3217},
archivePrefix = {arXiv},
       eprint = {2308.15735},
 primaryClass = {astro-ph.CO},
       adsurl = {https://ui.adsabs.harvard.edu/abs/2024MNRAS.527..501B}
}

@ARTICLE{gorski1988,
       author = {{Gorski}, Krzysztof},
        title = "{On the Pattern of Perturbations of the Hubble Flow}",
      journal = {\apjl},
         year = 1988,
        month = sep,
       volume = {332},
        pages = {L7},
          doi = {10.1086/185255},
       adsurl = {https://ui.adsabs.harvard.edu/abs/1988ApJ...332L...7G}
}

@ARTICLE{turner2021,
       author = {{Turner}, Ryan J. and {Blake}, Chris and {Ruggeri}, Rossana},
        title = "{Improving estimates of the growth rate using galaxy-velocity correlations: a simulation study}",
      journal = {\mnras},
         year = 2021,
        month = apr,
       volume = {502},
       number = {2},
        pages = {2087-2096},
          doi = {10.1093/mnras/stab212},
archivePrefix = {arXiv},
       eprint = {2101.09026},
 primaryClass = {astro-ph.CO},
       adsurl = {https://ui.adsabs.harvard.edu/abs/2021MNRAS.502.2087T}
}

@ARTICLE{1994ApJ...426...23F,
       author = {{Feldman}, Hume A. and {Kaiser}, Nick and {Peacock}, John A.},
        title = "{Power-Spectrum Analysis of Three-dimensional Redshift Surveys}",
      journal = {\apj},
         year = 1994,
        month = may,
       volume = {426},
        pages = {23},
          doi = {10.1086/174036},
archivePrefix = {arXiv},
       eprint = {astro-ph/9304022},
 primaryClass = {astro-ph},
       adsurl = {https://ui.adsabs.harvard.edu/abs/1994ApJ...426...23F}
}

@ARTICLE{landy1993,
       author = {{Landy}, Stephen D. and {Szalay}, Alexander S.},
        title = "{Bias and Variance of Angular Correlation Functions}",
      journal = {\apj},
         year = 1993,
        month = jul,
       volume = {412},
        pages = {64},
          doi = {10.1086/172900},
       adsurl = {https://ui.adsabs.harvard.edu/abs/1993ApJ...412...64L}
}

@ARTICLE{2019PASP..131a8002B,
       author = {{Bellm}, Eric C. and {Kulkarni}, Shrinivas R. and {Graham}, Matthew J. and {Dekany}, Richard and {Smith}, Roger M. and {Riddle}, Reed and {Masci}, Frank J. and {Helou}, George and {Prince}, Thomas A. and {Adams}, Scott M. and {Barbarino}, C. and {Barlow}, Tom and {Bauer}, James and {Beck}, Ron and {Belicki}, Justin and {Biswas}, Rahul and {Blagorodnova}, Nadejda and {Bodewits}, Dennis and {Bolin}, Bryce and {Brinnel}, Valery and {Brooke}, Tim and {Bue}, Brian and {Bulla}, Mattia and {Burruss}, Rick and {Cenko}, S. Bradley and {Chang}, Chan-Kao and {Connolly}, Andrew and {Coughlin}, Michael and {Cromer}, John and {Cunningham}, Virginia and {De}, Kishalay and {Delacroix}, Alex and {Desai}, Vandana and {Duev}, Dmitry A. and {Eadie}, Gwendolyn and {Farnham}, Tony L. and {Feeney}, Michael and {Feindt}, Ulrich and {Flynn}, David and {Franckowiak}, Anna and {Frederick}, S. and {Fremling}, C. and {Gal-Yam}, Avishay and {Gezari}, Suvi and {Giomi}, Matteo and {Goldstein}, Daniel A. and {Golkhou}, V. Zach and {Goobar}, Ariel and {Groom}, Steven and {Hacopians}, Eugean and {Hale}, David and {Henning}, John and {Ho}, Anna Y.~Q. and {Hover}, David and {Howell}, Justin and {Hung}, Tiara and {Huppenkothen}, Daniela and {Imel}, David and {Ip}, Wing-Huen and {Ivezi{\'c}}, {\v{Z}}eljko and {Jackson}, Edward and {Jones}, Lynne and {Juric}, Mario and {Kasliwal}, Mansi M. and {Kaspi}, S. and {Kaye}, Stephen and {Kelley}, Michael S.~P. and {Kowalski}, Marek and {Kramer}, Emily and {Kupfer}, Thomas and {Landry}, Walter and {Laher}, Russ R. and {Lee}, Chien-De and {Lin}, Hsing Wen and {Lin}, Zhong-Yi and {Lunnan}, Ragnhild and {Giomi}, Matteo and {Mahabal}, Ashish and {Mao}, Peter and {Miller}, Adam A. and {Monkewitz}, Serge and {Murphy}, Patrick and {Ngeow}, Chow-Choong and {Nordin}, Jakob and {Nugent}, Peter and {Ofek}, Eran and {Patterson}, Maria T. and {Penprase}, Bryan and {Porter}, Michael and {Rauch}, Ludwig and {Rebbapragada}, Umaa and {Reiley}, Dan and {Rigault}, Mickael and {Rodriguez}, Hector and {van Roestel}, Jan and {Rusholme}, Ben and {van Santen}, Jakob and {Schulze}, S. and {Shupe}, David L. and {Singer}, Leo P. and {Soumagnac}, Maayane T. and {Stein}, Robert and {Surace}, Jason and {Sollerman}, Jesper and {Szkody}, Paula and {Taddia}, F. and {Terek}, Scott and {Van Sistine}, Angela and {van Velzen}, Sjoert and {Vestrand}, W. Thomas and {Walters}, Richard and {Ward}, Charlotte and {Ye}, Quan-Zhi and {Yu}, Po-Chieh and {Yan}, Lin and {Zolkower}, Jeffry},
        title = "{The Zwicky Transient Facility: System Overview, Performance, and First Results}",
      journal = {\pasp},
         year = 2019,
        month = jan,
       volume = {131},
       number = {995},
        pages = {018002},
          doi = {10.1088/1538-3873/aaecbe},
archivePrefix = {arXiv},
       eprint = {1902.01932},
 primaryClass = {astro-ph.IM},
       adsurl = {https://ui.adsabs.harvard.edu/abs/2019PASP..131a8002B}
}

@ARTICLE{2020PDU....2900519G,
       author = {{Garcia}, Karolina and {Quartin}, Miguel and {Siffert}, Beatriz B.},
        title = "{On the amount of peculiar velocity field information in supernovae from LSST and beyond}",
      journal = {Physics of the Dark Universe},
         year = 2020,
        month = sep,
       volume = {29},
          eid = {100519},
        pages = {100519},
          doi = {10.1016/j.dark.2020.100519},
archivePrefix = {arXiv},
       eprint = {1905.00746},
 primaryClass = {astro-ph.CO},
       adsurl = {https://ui.adsabs.harvard.edu/abs/2020PDU....2900519G}
}

@ARTICLE{2023A&A...674A.197C,
       author = {{Carreres}, Bastien and {Bautista}, Julian E. and {Feinstein}, Fabrice and {Fouchez}, Dominique and {Racine}, Benjamin and {Smith}, Mathew and {Amenouche}, Melissa and {Aubert}, Marie and {Dhawan}, Suhail and {Ginolin}, Madeleine and {Goobar}, Ariel and {Gris}, Philippe and {Lacroix}, Leander and {Nuss}, Eric and {Regnault}, Nicolas and {Rigault}, Mickael and {Robert}, Estelle and {Rosnet}, Philippe and {Sommer}, Kelian and {Dekany}, Richard and {Groom}, Steven L. and {Sravan}, Niharika and {Masci}, Frank J. and {Purdum}, Josiah},
        title = "{Growth-rate measurement with type-Ia supernovae using ZTF survey simulations}",
      journal = {\aap},
         year = 2023,
        month = jun,
       volume = {674},
          eid = {A197},
        pages = {A197},
          doi = {10.1051/0004-6361/202346173},
archivePrefix = {arXiv},
       eprint = {2303.01198},
 primaryClass = {astro-ph.CO},
       adsurl = {https://ui.adsabs.harvard.edu/abs/2023A&A...674A.197C}
}

@ARTICLE{strauss1995,
       author = {{Strauss}, M.~A. and {Willick}, J.~A.},
        title = "{The density and peculiar velocity fields of nearby galaxies}",
      journal = {\physrep},
         year = 1995,
        month = jan,
       volume = {261},
        pages = {271-431},
          doi = {10.1016/0370-1573(95)00013-7},
archivePrefix = {arXiv},
       eprint = {astro-ph/9502079},
 primaryClass = {astro-ph},
       adsurl = {https://ui.adsabs.harvard.edu/abs/1995PhR...261..271S}
}

@ARTICLE{johnson2014,
       author = {{Johnson}, Andrew and {Blake}, Chris and {Koda}, Jun and {Ma}, Yin-Zhe and {Colless}, Matthew and {Crocce}, Martin and {Davis}, Tamara M. and {Jones}, Heath and {Magoulas}, Christina and {Lucey}, John R. and {Mould}, Jeremy and {Scrimgeour}, Morag I. and {Springob}, Christopher M.},
        title = "{The 6dF Galaxy Survey: cosmological constraints from the velocity power spectrum}",
      journal = {\mnras},
         year = 2014,
        month = nov,
       volume = {444},
       number = {4},
        pages = {3926-3947},
          doi = {10.1093/mnras/stu1615},
archivePrefix = {arXiv},
       eprint = {1404.3799},
 primaryClass = {astro-ph.CO},
       adsurl = {https://ui.adsabs.harvard.edu/abs/2014MNRAS.444.3926J}
}

@ARTICLE{huterer2017,
       author = {{Huterer}, Dragan and {Shafer}, Daniel L. and {Scolnic}, Daniel M. and {Schmidt}, Fabian},
        title = "{Testing {\ensuremath{\Lambda}}CDM at the lowest redshifts with SN Ia and galaxy velocities}",
      journal = {\jcap},
         year = 2017,
        month = may,
       volume = {2017},
       number = {5},
          eid = {015},
        pages = {015},
          doi = {10.1088/1475-7516/2017/05/015},
archivePrefix = {arXiv},
       eprint = {1611.09862},
 primaryClass = {astro-ph.CO},
       adsurl = {https://ui.adsabs.harvard.edu/abs/2017JCAP...05..015H}
}

@ARTICLE{howlett2017b,
       author = {{Howlett}, Cullan and {Staveley-Smith}, Lister and {Elahi}, Pascal J. and {Hong}, Tao and {Jarrett}, Tom H. and {Jones}, D. Heath and {Koribalski}, B{\"a}rbel S. and {Macri}, Lucas M. and {Masters}, Karen L. and {Springob}, Christopher M.},
        title = "{2MTF - VI. Measuring the velocity power spectrum}",
      journal = {\mnras},
         year = 2017,
        month = nov,
       volume = {471},
       number = {3},
        pages = {3135-3151},
          doi = {10.1093/mnras/stx1521},
archivePrefix = {arXiv},
       eprint = {1706.05130},
 primaryClass = {astro-ph.CO},
       adsurl = {https://ui.adsabs.harvard.edu/abs/2017MNRAS.471.3135H}
}

@ARTICLE{adams2020,
       author = {{Adams}, Caitlin and {Blake}, Chris},
        title = "{Joint growth-rate measurements from redshift-space distortions and peculiar velocities in the 6dF Galaxy Survey}",
      journal = {\mnras},
         year = 2020,
        month = may,
       volume = {494},
       number = {3},
        pages = {3275-3293},
          doi = {10.1093/mnras/staa845},
archivePrefix = {arXiv},
       eprint = {2004.06399},
 primaryClass = {astro-ph.CO},
       adsurl = {https://ui.adsabs.harvard.edu/abs/2020MNRAS.494.3275A}
}

@ARTICLE{lai2023,
       author = {{Lai}, Yan and {Howlett}, Cullan and {Davis}, Tamara M.},
        title = "{Using peculiar velocity surveys to constrain the growth rate of structure with the wide-angle effect}",
      journal = {\mnras},
         year = 2023,
        month = jan,
       volume = {518},
       number = {2},
        pages = {1840-1858},
          doi = {10.1093/mnras/stac3252},
archivePrefix = {arXiv},
       eprint = {2209.04166},
 primaryClass = {astro-ph.CO},
       adsurl = {https://ui.adsabs.harvard.edu/abs/2023MNRAS.518.1840L}
}

\section*{Affiliations}
\scriptsize
\noindent
$^{1}$ Centre for Astrophysics \& Supercomputing, Swinburne University of Technology, P.O. Box 218, Hawthorn, VIC 3122, Australia\\
$^{2}$ OzGrav: The ARC Centre of Excellence for Gravitational Wave Discovery\\
$^{3}$ Aix Marseille Univ, CNRS/IN2P3, CPPM, Marseille, France\\
$^{4}$ Lawrence Berkeley National Laboratory, 1 Cyclotron Road, Berkeley, CA 94720, USA\\
$^{5}$ Department of Physics, Boston University, 590 Commonwealth Avenue, Boston, MA 02215 USA\\
$^{6}$ Department of Physics \& Astronomy, University of Rochester, 206 Bausch and Lomb Hall, P.O. Box 270171, Rochester, NY 14627-0171, USA\\
$^{7}$ Dipartimento di Fisica ``Aldo Pontremoli'', Universit\`a degli Studi di Milano, Via Celoria 16, I-20133 Milano, Italy\\
$^{8}$ INAF-Osservatorio Astronomico di Brera, Via Brera 28, 20122 Milano, Italy\\
$^{9}$ Department of Physics \& Astronomy, University College London, Gower Street, London, WC1E 6BT, UK\\
$^{10}$ Korea Astronomy and Space Science Institute, 776, Daedeokdae-ro, Yuseong-gu, Daejeon 34055, Republic of Korea\\
$^{11}$ Instituto de F\'{\i}sica, Universidad Nacional Aut\'{o}noma de M\'{e}xico,  Circuito de la Investigaci\'{o}n Cient\'{\i}fica, Ciudad Universitaria, Cd. de M\'{e}xico  C.~P.~04510,  M\'{e}xico\\
$^{12}$ Department of Astronomy \& Astrophysics, University of Toronto, Toronto, ON M5S 3H4, Canada\\
$^{13}$ Department of Physics \& Astronomy and Pittsburgh Particle Physics, Astrophysics, and Cosmology Center (PITT PACC), University of Pittsburgh, 3941 O'Hara Street, Pittsburgh, PA 15260, USA\\
$^{14}$ University of California, Berkeley, 110 Sproul Hall \#5800 Berkeley, CA 94720, USA\\
$^{15}$ Departamento de F\'isica, Universidad de los Andes, Cra. 1 No. 18A-10, Edificio Ip, CP 111711, Bogot\'a, Colombia\\
$^{16}$ Observatorio Astron\'omico, Universidad de los Andes, Cra. 1 No. 18A-10, Edificio H, CP 111711 Bogot\'a, Colombia\\
$^{17}$ Institut d'Estudis Espacials de Catalunya (IEEC), c/ Esteve Terradas 1, Edifici RDIT, Campus PMT-UPC, 08860 Castelldefels, Spain\\
$^{18}$ Institute of Cosmology and Gravitation, University of Portsmouth, Dennis Sciama Building, Portsmouth, PO1 3FX, UK\\
$^{19}$ Institute of Space Sciences, ICE-CSIC, Campus UAB, Carrer de Can Magrans s/n, 08913 Bellaterra, Barcelona, Spain\\
$^{20}$ University of Virginia, Department of Astronomy, Charlottesville, VA 22904, USA\\
$^{21}$ Fermi National Accelerator Laboratory, PO Box 500, Batavia, IL 60510, USA\\
$^{22}$ Center for Cosmology and AstroParticle Physics, The Ohio State University, 191 West Woodruff Avenue, Columbus, OH 43210, USA\\
$^{23}$ Department of Physics, The Ohio State University, 191 West Woodruff Avenue, Columbus, OH 43210, USA\\
$^{24}$ The Ohio State University, Columbus, 43210 OH, USA\\
$^{25}$ School of Mathematics and Physics, University of Queensland, Brisbane, QLD 4072, Australia\\
$^{26}$ Department of Physics, University of Michigan, 450 Church Street, Ann Arbor, MI 48109, USA\\
$^{27}$ University of Michigan, 500 S. State Street, Ann Arbor, MI 48109, USA\\
$^{28}$ Department of Physics, The University of Texas at Dallas, 800 W. Campbell Rd., Richardson, TX 75080, USA\\
$^{29}$ NSF NOIRLab, 950 N. Cherry Ave., Tucson, AZ 85719, USA\\
$^{30}$ Department of Physics, Southern Methodist University, 3215 Daniel Avenue, Dallas, TX 75275, USA\\
$^{31}$ Sorbonne Universit\'{e}, CNRS/IN2P3, Laboratoire de Physique Nucl\'{e}aire et de Hautes Energies (LPNHE), FR-75005 Paris, France\\
$^{32}$ Department of Astronomy and Astrophysics, UCO/Lick Observatory, University of California, 1156 High Street, Santa Cruz, CA 95064, USA\\
$^{33}$ Department of Astronomy and Astrophysics, University of California, Santa Cruz, 1156 High Street, Santa Cruz, CA 95065, USA\\
$^{34}$ Departament de F\'{i}sica, Serra H\'{u}nter, Universitat Aut\`{o}noma de Barcelona, 08193 Bellaterra (Barcelona), Spain\\
$^{35}$ Institut de F\'{i}sica d’Altes Energies (IFAE), The Barcelona Institute of Science and Technology, Edifici Cn, Campus UAB, 08193, Bellaterra (Barcelona), Spain\\
$^{36}$ Department of Astronomy, The Ohio State University, 4055 McPherson Laboratory, 140 W 18th Avenue, Columbus, OH 43210, USA\\
$^{37}$ Instituci\'{o} Catalana de Recerca i Estudis Avan\c{c}ats, Passeig de Llu\'{\i}s Companys, 23, 08010 Barcelona, Spain\\
$^{38}$ Department of Physics and Astronomy, University of Sussex, Brighton BN1 9QH, U.K\\
$^{39}$ IRFU, CEA, Universit\'{e} Paris-Saclay, F-91191 Gif-sur-Yvette, France\\
$^{40}$ Department of Physics and Astronomy, University of Waterloo, 200 University Ave W, Waterloo, ON N2L 3G1, Canada\\
$^{41}$ Perimeter Institute for Theoretical Physics, 31 Caroline St. North, Waterloo, ON N2L 2Y5, Canada\\
$^{42}$ Waterloo Centre for Astrophysics, University of Waterloo, 200 University Ave W, Waterloo, ON N2L 3G1, Canada\\
$^{43}$ Space Sciences Laboratory, University of California, Berkeley, 7 Gauss Way, Berkeley, CA  94720, USA\\
$^{44}$ Instituto de Astrof\'{i}sica de Andaluc\'{i}a (CSIC), Glorieta de la Astronom\'{i}a, s/n, E-18008 Granada, Spain\\
$^{45}$ Department of Physics and Astronomy, Sejong University, 209 Neungdong-ro, Gwangjin-gu, Seoul 05006, Republic of Korea\\
$^{46}$ CIEMAT, Avenida Complutense 40, E-28040 Madrid, Spain\\
$^{47}$ National Astronomical Observatories, Chinese Academy of Sciences, A20 Datun Road, Chaoyang District, Beijing, 100101, P.~R.~China\\
\normalsize

\end{document}